\documentclass[aps,prc,preprint,showpacs,preprintnumbers,
                          nofootinbib,float,superscriptaddress,longbibliography]{revtex4-1}
\usepackage{graphicx, fancybox}
\usepackage{amsmath,amssymb}
\usepackage[colorlinks=true, pdfstartview=FitV, linkcolor=red,
		     citecolor=blue, urlcolor=blue]{hyperref}
\usepackage{color}
\usepackage{soul}
\usepackage{url}
\usepackage[normalem]{ulem}

\begin{document}

\title{Charge-dependent Flow Induced by Magnetic and Electric Fields in Heavy Ion Collisions}

\author{Umut G\"ursoy}
\affiliation{Institute for Theoretical Physics, Utrecht University, Leuvenlaan 4, 3584 CE Utrecht, The Netherlands}

\author{Dmitri Kharzeev}
\affiliation{Department of Physics and Astronomy, Stony Brook University, New York 11794, USA}
\affiliation{Physics Department and RIKEN-BNL Research Center, Brookhaven National Laboratory, Upton, NY 11973, USA}

\author{Eric Marcus}
\affiliation{Institute for Theoretical Physics, Utrecht University, Leuvenlaan 4, 3584 CE Utrecht, The Netherlands}

\author{Krishna Rajagopal}
\affiliation{Center for Theoretical Physics,\\ Massachusetts Institute of Technology, Cambridge, MA 02139}

\author{Chun Shen}
\affiliation{Physics Department, Brookhaven National Laboratory, Upton, NY 11973, USA}

\preprint{MIT-CTP-5012}

\begin{abstract}
We investigate the charge-dependent flow induced by magnetic and electric fields
in heavy ion collisions. We simulate the evolution of the expanding cooling droplet of strongly coupled plasma hydrodynamically, using
the {\tt iEBE-VISHNU} framework, and add the magnetic and electric fields as well as the electric currents they generate in a perturbative fashion.  
We confirm the previously reported effect of the electromagnetically induced currents~\cite{Gursoy:2014aka}, that is a charge-odd directed flow
$\Delta v_1$ that is odd in rapidity, noting that it is induced by magnetic fields (\`a la Faraday and Lorentz) and by electric fields (the Coulomb field from the charged spectators).
In addition, we find a charge-odd $\Delta v_3$ that is also odd in rapidity and that has a similar physical origin.
We furthermore show that the electric field  produced by the net charge density of the plasma drives rapidity-even charge-dependent contributions to the radial flow $\langle p_T \rangle$ and the elliptic flow $\Delta v_2$. Although their magnitudes are comparable to the charge-odd $\Delta v_1$ and $\Delta v_3$, they have a different physical origin, namely the Coulomb forces within the plasma.  
\end{abstract}

\maketitle
\date{\today}

\section{Introduction}
\label{sec::intro}

Large magnetic fields $\vec B$ are produced 
in all non-central 
heavy ion collisions (those with nonzero impact parameter) by the moving and positively charged spectator nucleons that ``miss'', flying past each other rather than colliding, as well as by the nucleons that participate in the collision. Estimates obtained by applying the Biot-Savart law to collisions with an
impact parameter $b=4$~fm yield $e|\vec B|/m_\pi^2 \approx$ 1-3 about 0.1-0.2 fm$/c$
after a RHIC collision with $\sqrt{s}=200$~AGeV 
and $e|\vec B|/
m_\pi^2 \approx $ 10-15 at some even earlier time after 
an LHC collision with $\sqrt{s}=2.76$~ATeV~\cite{Kharzeev:2007jp,Skokov:2009qp,Tuchin:2010vs,Voronyuk:2011jd,Deng:2012pc,Tuchin:2013ie,McLerran:2013hla,Gursoy:2014aka}. 
The interplay between these magnetic fields and quantum anomalies
has been of much interest in recent years, as it has been predicted to 
lead to interesting phenomena including the chiral magnetic effect~\cite{Kharzeev:2007jp,Fukushima:2008xe}
and the chiral magnetic wave~\cite{Kharzeev:2010gd,Burnier:2011bf}.
This makes it imperative to establish that the presence of an early-time magnetic
field can, via Faraday's Law and the Lorentz force, 
have observable consequences on the motion of the final-state charged
particles seen in the detectors~\cite{Gursoy:2014aka}.
Since the plasma produced in collisions of positively charged nuclei has a (small) net positive charge, electric effects -- which is to say the Coulomb force -- can also yield observable consequences to the motion of charged particles in the final state. These electric effects are distinct from the consequences of a magnetic field first studied in Ref.~\cite{Gursoy:2014aka}, but comparable in magnitude.
Our goal in this paper will be a qualitative, perhaps semi-quantitative, assessment
of the observable effects of both magnetic and electric fields, arising just via the Maxwell equations
and the Lorentz force law,
so that experimental measurements can be used to constrain the 
strength of the fields 
and to establish baseline expectations against which to compare any other, possibly anomalous, experimental consequences of $\vec B$.   

In previous work \cite{Gursoy:2014aka} three of the authors noted that the magnetic field produced
in a heavy ion collision
could result in a measurable effect in the form of a charge-odd contribution to the directed flow coefficient $\Delta v_1$. This contribution has the opposite sign for positively vs. negatively charged hadrons in the final state and is odd in rapidity. However, the authors of  \cite{Gursoy:2014aka} neglected to observe that a part of this charge-odd, parity-odd effect originates from the Coulomb interaction. In particular it originates from   the interaction between the positively charged 
spectators that have passed by the collision and the plasma produced in the collision, as will be explained in detail below.

The study in Ref.~\cite{Gursoy:2014aka} was simplified in many ways, including in particular by being built upon the azimuthally symmetric 
solution to the equations of relativistic viscous hydrodynamics
constructed by Gubser 
in Ref.~\cite{Gubser:2010ze}. Because this solution is analytic, various practical simplifications in the calculations of Ref.~\cite{Gursoy:2014aka} followed. 
In reality magnetic fields do not arise in azimuthally symmetric collisions.
The calculations of Ref.~\cite{Gursoy:2014aka} were intended to provide an initial order of magnitude estimate of the $\vec B$-driven, charge-odd, rapidity-odd contribution to $\Delta v_1$ in heavy ion collisions with a nonzero impact parameter, but the authors perturbed around an azimuthally symmetric hydrodynamic solution for simplicity.
Also, the radial profile of the energy density in Gubser's solution to hydrodynamics is not realistic.  
Here, we shall repeat and extend the calculation of Ref.~\cite{Gursoy:2014aka}, this time building
the perturbative calculation of the electromagnetic fields 
and the resulting currents upon 
numerical solutions to the equations of relativistic viscous hydrodynamics 
simulated within the {\tt iEBE-VISHNU} framework~\cite{Shen:2014vra} that provide a good 
description of azimuthally anisotropic heavy ion collisions with a nonzero impact parameter.

\begin{figure}[t]
\vspace{-0.2in}
 \begin{center}
\includegraphics[scale=0.45]{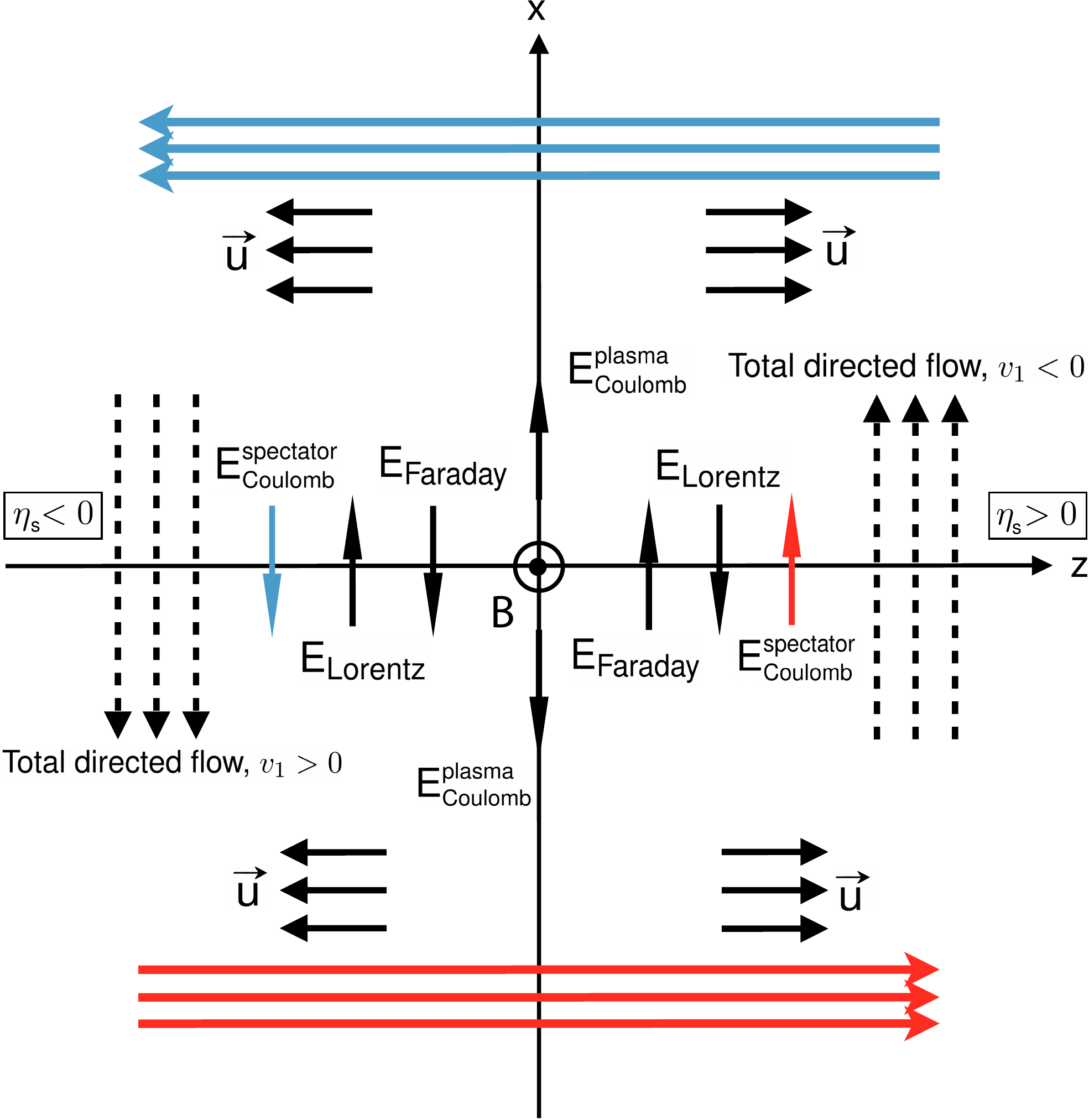}
 \end{center}
 \vspace{-0.3in}
 \caption{Schematic illustration of how the magnetic field $\vec B$ in a heavy ion collision results
in a directed flow {\it of electric charge}, $\Delta v_1$.  The collision occurs in the $z$-direction, meaning that
the longitudinal expansion velocity $\vec u$ of the conducting QGP produced in
the collision points in the $+z$ ($-z$)
direction at positive (negative) $z$.  We take the impact parameter vector to point in the $+x$ direction,
choosing the nucleus moving toward positive (negative) $z$ to be located at  negative (positive) $x$. 
The trajectories of the spectators that ``miss'' the collision because of the nonzero impact parameter are indicated by the red and blue arrows.
This configuration generates a magnetic field $\vec B$ in the $+y$ direction, as shown.
The directions of the electric fields (and hence currents)
due to the Faraday, Lorentz and Coulomb effects are shown. The
two different Coulomb contributions are indicated, one due to the force exerted by the spectators and the other coming from Coulomb forces within the plasma.
The dashed arrows indicate the
direction of the directed flow of positive charge in the case where
the Faraday + spectator Coulomb effects are on balance stronger than the Lorentz effect. 
Hence, the total directed flow in this example corresponds to $v_1 < 0$ ($v_1>0$) for positive charges at spacetime rapidity $\eta_s>0$ ($\eta_s<0$), and opposite for negative charges.}
\label{figschema}
\end{figure}

The idea of Ref.~\cite{Gursoy:2014aka} is to calculate the electromagnetic fields, and then the incremental contribution to the velocity fields of the positively and negatively charged components of the 
hydrodynamic fluid (aka the electric currents) caused by the electromagnetic forces, in a perturbative manner. A similar 
conclusion have been reached in \cite{Roy:2017yvg} and  \cite{Stewart:2017zsu}. 
One first computes the electric and magnetic fields $\vec E$ and $\vec B$ using the Maxwell equations as we describe further below.
Then, at each point in the fluid, one
transforms to the local fluid rest frame by boosting with the local 
background velocity field $\vec{v}_\mathrm{flow}$. Afterwards one computes the incremental drift velocity $\vec{v}_\mathrm{drift}$
caused by the electromagnetic forces in this frame
by demanding that the electromagnetic force acting on a fluid unit cell 
with charge $q$ is balanced by the drag force. One then boosts 
back to the lab frame to obtain the total velocity field that now includes 
both $\vec{v}_\mathrm{flow}$ and $\vec{v}_\mathrm{drift}$, with 
$\vec{v}_\mathrm{drift}$ taking opposite signs for the positively and negatively charged
components of the fluid. The authors of Ref.~\cite{Gursoy:2014aka} then use
a standard Cooper-Frye freeze out analysis 
to show that the electromagnetic forces acting within the hydrodynamic fluid result in 
a contribution to the charge-odd
directed flow parameter $\Delta v_1 \equiv v_1(h^+) - v_1(h^-)$. We shall provide the (standard) definition of the directed flow $v_1$ in Section~\ref{sec::model}. The charge-odd contribution $\Delta v_1$ is small but distinctive:
in addition to being anti-symmetric under the flip of charge, it is also antisymmetric under flipping the rapidity. 
That the contribution has opposite sign for oppositely charged hadrons is easy to understand: it results from an electric current in the plasma.
The fact that it has opposite sign at positive and negative rapidity can also easily be understood, as we explain in  Fig.~\ref{figschema} and below.

As illustrated in Fig.~\ref{figschema}, there are three distinct origins for 
a sideways push on charged components of the fluid, resulting in a sideways current:
\begin{enumerate}
\item {\it Faraday:} as the magnetic field decreases in time (see the right panel of  Fig.~\ref{figEeta0} below), Faraday's law dictates the induction of an electric field and, since the plasma includes mobile charges, an electric current. We denote this electric field by $\vec E_F$.  Since $\vec E_F$ curls around the (decreasing) $\vec B$ that points in the $y$-direction, the sideways component of $E_F$ points in opposite directions at opposite rapidity, see Fig.~\ref{figschema}.
\item {\it Lorentz:} since the hydrodynamic fluid exhibits a strong longitudinal flow velocity $\vec{v}_\mathrm{flow}$ denoted by $\vec u$ in Fig.~\ref{figschema}, which points along the beam direction (hence perpendicular to $\vec B$), the Lorentz force exerts a sideways push on charged particles in opposite directions at opposite rapidity. Equivalently, upon boosting to the local fluid rest frame in which the fluid is not moving, the lab frame $\vec B$ yields a fluid frame $\vec E$ whose effects on the charged components of the fluid are equivalent to the effects of the Lorentz force in the lab frame. We denote this electric field by $\vec E_L$.  Both $\vec E_F$ and $\vec E_L$ are of magnetic origin.
\item {\it Coulomb:} The positively charged spectators that have passed the collision zone exert an electric force on the charged plasma produced in the collision, which again points in opposite directions at opposite rapidity.
We denote this electric field by $\vec E_C$. As we noted above, the authors of Ref.~\cite{Gursoy:2014aka} did not 
identify this contribution, even though it was correctly included in their numerical results.
\end{enumerate}
As is clear from their physical origins, all three of these electric fields --- and the consequent electric currents --- have opposite directions at positive and negative rapidity. It is also clear from Fig.~\ref{figschema} 
that $\vec E_F$ and  $\vec E_C$ have the same sign, while  $\vec E_L$ opposes them. Hence,
 the sign of the total rapidity-odd, charge-odd, $\Delta v_1$ that results 
 from the electric current driven by these electric fields
 depends on whether  $\vec E_F + \vec E_C$ or $\vec E_L$ is dominant.

In this paper we make three significant advances relative to the exploratory study of Ref.~\cite{Gursoy:2014aka}.  First, as already noted we build our calculation upon a realistic hydrodynamic description of the expansion dynamics of the droplet of matter produced in a heavy ion collision with a nonzero impact parameter. 

Second, we find that the same mechanism that produces the charge-odd $\Delta v_1$ also produces a similar charge-odd contribution to all the odd flow coefficients.  The azimuthal asymmetry of the almond-shaped collision zone in a collision with nonzero impact parameter, its remaining symmetries under  $x \leftrightarrow -x$ and $y\leftrightarrow -y$, and the orientation of the magnetic field $\vec B$ perpendicular to the beam and impact parameter directions together mean that the currents induced by the Faraday and Lorentz effects (illustrated in
Fig.~\ref{figschema}) make a charge-odd and rapidity-odd contribution
to all the odd flow harmonics, not only to $\Delta v_1$.  We compute the charge-odd contribution to $\Delta v_3$ in addition to $\Delta v_1$ in this paper.

Last but not least, we identify a new electromagnetic mechanism that generates another type of sideways current which generates a charge-odd, {\it rapidity-even}, contribution to the {\it elliptical} flow coefficient $\Delta v_2$. Although it differs in its symmetry from the three sources of sideways electric field above, it should be added to our list:
\begin{enumerate}
\item[4.] {\it Plasma:} As is apparent from the left panel of Fig.~\ref{figEB} in Section~\ref{sec::EM} and as we show explicitly in that Section, there is a non-vanishing outward-pointing component of the electric field already in the lab frame, because the plasma (and the spectators) have a net positive charge. We denote this component of the electric field by $\vec E_P$, since its origin includes Coulomb forces within the plasma.  
\end{enumerate} 
At the collision energies that we consider, $\vec E_P$ receives contributions both from the spectator nucleons and from
the charge density deposited in the plasma by the nucleons participating in the collision.
As illustrated below by the results in the left panel of Fig.~\ref{figEB}, the electric field will push an outward-directed current.  
As this field configuration is even in rapidity and odd under $x\leftrightarrow -x$ (which means that the radial component of the field is even under $x\leftrightarrow -x$), the current that it drives will yield a  {\it rapidity-even}, charge-odd, contribution to
the {\it even} flow harmonics, see Fig.~\ref{figschema}.  
We shall demonstrate this by calculating the
charge-dependent contribution to the radial flow, $\Delta \langle p_T \rangle$ (which can be thought of as $\Delta v_0$) and to the elliptic flow, $\Delta v_2$, that result from the electric field $\vec E_P$. 
Furthermore, we discover that these observables also receive a contribution from a component of the spectator-induced contribution to the electric field $\vec E_F + \vec E_L + \vec E_C$ that is odd under $x\leftrightarrow -x$ and even in rapidity.

In the next Section, we set up our model. In particular, we explain our calculation of the electromagnetic fields, the drift velocity and the freezeout procedure from which we read off the charge-dependent contributions to the radial $\langle p_T\rangle$ and to the anisotropic flow parameters $v_1$, $v_2$ and $v_3$. In Section \ref{sec::EM} we present numerical results for the electromagnetic fields. Then in Section~\ref{sec::results} we move on to the calculation of the flow coefficients, for  collisions with both RHIC and LHC energies, for pions and for protons, for varying centralities and ranges of $p_T$, and for several values of the electrical conductivity $\sigma$ of the plasma and the drag coefficient $\mu m$. The latter two being the properties of the plasma to which the effects that we analyse are sensitive. Finally in Section~\ref{sec::discussion} we discuss the validity of the various approximations used in our calculations, discuss other related work, and present an outlook.

\section{Model Setup}
\label{sec::model}

We simulate the dynamical evolution of the medium produced in heavy-ion collisions using the {\tt iEBE-VISHNU} framework described in full in Ref.~\cite{Shen:2014vra}. We take event-averaged initial conditions from a Monte-Carlo-Glauber model, obtaining the initial energy density profiles 
by first aligning individual bumpy events with respect to their second-order participant plane angles (the appropriate proxy for the reaction plane in a bumpy event) and then averaging over 10,000 events. 
The second order participant plane of the averaged initial condition, $\Psi_2^\mathrm{PP}$,  is rotated to align with the $x$-axis, which is to say we choose coordinates such that the averaged initial condition has $\Psi_2^\mathrm{PP} = 0$ and an impact parameter vector that points in the $+x$ direction. The 
hydrodynamic calculation that follows
assumes longitudinal boost-invariance and starts at $\tau_0 = 0.4$ fm/$c$.\footnote{Starting hydrodynamics at a different thermalization time, between 0.2 and 0.6 fm/$c$, only changes the hadronic observables by few percent.\cite{Shen:2010uy} } We then evolve the relativistic viscous
hydrodynamic equations for a fluid with an equation of state based upon lattice QCD calculations, choosing the s95p-v1-PCE equation of state from Ref.~\cite{Huovinen:2009yb} which implements partial chemical equilibrium at $T_\mathrm{chem} = 150$ MeV. 
The kinetic freeze-out temperature is fixed to be 105 MeV to reproduce the mean $p_T$ of the identified hadrons in the final state.
Specifying the equations of relativistic viscous hydrodynamics requires specifying the temperature dependent ratio of the shear viscosity to the entropy density, $\eta/s(T)$, in addition to specifying the equation of state.  Following Ref.~\cite{Niemi:2011ix}, we choose
\begin{equation}
\frac{\eta}{s} (T) = \left\{ \begin{array}{rcl} \left(\frac{\eta}{s}\right)_\mathrm{min} + 0.288 \left(\frac{T}{T_\mathrm{tr}} - 1 \right) + 0.0818 \left(\left(\frac{T}{T_\mathrm{tr}}\right)^2 - 1\right) & \mbox{for} & T > T_\mathrm{tr} \\ \left(\frac{\eta}{s}\right)_\mathrm{min} + 0.0594 \left(1 - \frac{T}{T_\mathrm{tr}}\right) + 0.544 \left(1 - \left(\frac{T}{T_\mathrm{tr}}\right)^2 \right) & \mbox{for} & T < T_\mathrm{tr}  \end{array} \right. .
\label{eq1}
\end{equation}
We choose $(\eta/s)_\mathrm{min} = 0.08$ at $T_\mathrm{tr} = 180$ MeV. These choices result in hydrodynamic simulations that yield reasonable agreement with the experimental measurements over all centrality and collision energies, see for example Fig.~\ref{fig1} in Section~\ref{sec::results} below.

The electromagnetic fields are generated by both the spectators and participant charged nucleons. The transverse distribution of the right-going ($+$) and left-going ($-$) charge density profiles $\rho^\pm_\mathrm{spectator}(\vec{x}_\perp)$ and $\rho^\pm_\mathrm{participant} (\vec{x}_\perp)$ are generated by averaging over 10,000 events using the same Monte-Carlo-Glauber model used to initialize the hydrodynamic calculation. The external charge and current sources for the electromagnetic fields are then given by
\begin{equation}
\rho_\mathrm{ext} (\vec{x}_\perp, \eta_s) = \rho^+_\mathrm{ext} (\vec{x}_\perp, \eta_s) + \rho^-_\mathrm{ext} (\vec{x}_\perp, \eta_s)
\end{equation}
\begin{equation}
\vec{J}_\mathrm{ext} (\vec{x}_\perp, \eta_s) = \vec{J}^+_\mathrm{ext} (\vec{x}_\perp, \eta_s) + \vec{J}^-_\mathrm{ext} (\vec{x}_\perp, \eta_s)
\end{equation}
with
\begin{eqnarray}
\rho^\pm_\mathrm{ext} (\vec{x}_\perp, \eta_s) &=& \rho^\pm_\mathrm{spectator} (\vec{x}_\perp) \delta(\eta_s \mp y_\mathrm{beam}) + \rho^\pm_\mathrm{participant} (\vec{x}_\perp) f^{\pm}(\eta_s)  \label{eq2} \\
\vec{J}^\pm_\mathrm{ext}(\vec{x}_\perp, \eta_s) &=& \vec{\beta}^\pm(\eta_s) \rho^\pm_\mathrm{ext}(\vec{x}_\perp, \eta_s)  \quad \mbox{with} \quad \vec{\beta}^\pm = (0, 0, \pm \tanh(\eta_s)).
\end{eqnarray}
Here we are making the  
Bjorken approximation: the space-time rapidities $\eta_s$ of the external charges are assumed equal to their rapidity. The spectators fly with the beam rapidity $y_\mathrm{beam}$ and the participant nucleons lose some rapidity in the collisions; their rapidity distribution in Eq.~(\ref{eq2}) is assumed to be~\cite{Kharzeev:1996sq,Kharzeev:2007jp,Gursoy:2014aka}  
\begin{equation}
f^{\pm}(y) = \frac{1}{4 \sinh(y_\mathrm{beam}/2)} e^{\pm y/2} \quad \mbox{for} \quad  -y_\mathrm{beam} < y < y_\mathrm{beam}.
\label{ParticipantRapidityDistribution}
\end{equation}

The electromagnetic fields generated by the charges and currents evolve according to the Maxwell equations
\begin{eqnarray}
(\nabla^2 - \partial_t^2 - \sigma \partial_t) \vec{B} &=& - \vec{\nabla} \times \vec{J}_\mathrm{ext} \label{eq7} \\
(\nabla^2 - \partial_t^2 - \sigma \partial_t) \vec{E} &=& \frac{1}{\epsilon} \vec{\nabla} \rho_\mathrm{ext} + \partial_t \vec{J}_\mathrm{ext}\,. \label{eq8}
\end{eqnarray}
Here $\sigma$ is the electrical conductivity of the QGP plasma.  As in Ref.~\cite{Gursoy:2014aka}, we shall make the significant simplifying assumption of treating $\sigma$ as if it were a constant.
We make this assumption only because it permits us to use a semi-analytic form for the evolution of the electromagnetic fields rather than having to solve Eqs.~(\ref{eq7}) and (\ref{eq8}) fully numerically. This simplification therefore significantly speeds up our calculations. In reality, $\sigma$ is certainly temperature dependent: just on dimensional grounds it is expected to be proportional to the temperature of the plasma, meaning that $\sigma$ should be a function of space and time as the plasma expands and flows hydrodynamically, with $\sigma$ decreasing as the plasma cools.
Furthermore, during the pre-equilibrium epoch $\sigma$ should rapidly increase from zero to its equilibrium value.  Taking all of this into consideration would require a full, numerical, magnetohydrodynamical analysis, something that we leave for the future.
Throughout most of this paper, we shall follow Ref.~\cite{Gursoy:2014aka} and set 
the electrical conductivity to the constant value $\sigma = 0.023$ fm$^{-1}$ which, 
according to the lattice QCD calculations in Refs.~\cite{Ding:2010ga,Francis:2011bt,Brandt:2012jc,Amato:2013naa,Kaczmarek:2013dya}, corresponds to $\sigma$ in three-flavor quark-gluon plasma at $T\sim 250$~MeV.
The numerical code that we have used to compute the evolution of the electromagnetic fields can be found at \url{https://github.com/chunshen1987/Heavy-ion_EM_fields}.

With the evolution of the electromagnetic fields in hand, the next step is to compute the drift velocity $\vec v_{\rm drift}$ that the electromagnetic field induces at each point on the freeze-out surface.
Because this drift velocity  is only a small perturbation compared to the background hydrodynamic flow velocity, $\vert \vec{v}_\mathrm{drift} \vert \ll \vert \vec{v}_\mathrm{flow} \vert$, we can obtain $\vec v_{\rm drift}$ by solving the force-balance  equation~\cite{Gursoy:2014aka}
\begin{equation}
m \frac{d \vec{v}^\mathrm{\, lrf}_\mathrm{drift}}{dt} = q \vec{v}^\mathrm{,\ lrf}_\mathrm{drift} \times \vec{B}^\mathrm{\, lrf} + q \vec{E}^\mathrm{\, lrf} - \mu m \vec{v}^\mathrm{\, lrf}_\mathrm{drift} = 0
\label{eq9}
\end{equation}
in its non-relativistic form in the local rest frame of the fluid cell of interest.
The last term in (\ref{eq9})  describes the drag force on a fluid element with mass $m$ on
which some external (in this case electromagnetic) force is being exerted, with $\mu$ the drag
coefficient.
The calculation of $\mu m$ for quark-gluon plasma in QCD remains an open question.
In the ${\cal N}=4$ supersymmetric Yang-Mills (SYM) theory plasma it should be accessible
via a holographic calculation. At present its value is known precisely only for heavy quarks
in $\mathcal{N} = 4$ SYM theory, in which~\cite{Herzog:2006gh,CasalderreySolana:2006rq,Gubser:2006bz}, 
\begin{equation}
\mu m = \frac{\pi \sqrt{\lambda}}{2} T^2
\label{eq10}
\end{equation}
with $\lambda\equiv g^2 N_c$  the 't-Hooft coupling, $g$ being the gauge coupling and $N_c$ the number of colors. For our purposes, throughout most of this paper we shall follow Ref.~\cite{Gursoy:2014aka} and use (\ref{eq9}) with $\lambda = 6\pi$. We investigate the consequences of
varying this choice in Section~\ref{sec::parameters}.
Finally, the drift velocity $\vec{v}^\mathrm{\,lrf}_\mathrm{drift}$ in every fluid cell along the freeze-out surface  is boosted by the flow velocity to bring it back  to the lab frame, $V^\mu = (\Lambda_\mathrm{flow})^\mu\,_\nu(u^\mathrm{lrf}_\mathrm{drift})^\nu$, where $(\Lambda_\mathrm{flow})^\mu\,_\nu$ is the Lorentz boost matrix associated with the hydrodynamic flow velocity $u^\mu_\mathrm{flow}$.

With the full, charge-dependent, fluid velocity $V^\mu$ --- including the sum of the flow velocity and the charge-dependent drift velocity induced by the electromagnetic fields ---  in hand, we now use the Cooper-Frye formula~\cite{Cooper:1974mv},
\begin{equation}
\frac{d N}{d y p_T d p_T d \phi} = \frac{g}{(2 \pi)^3} \int_\Sigma p^{\mu} d
  \sigma_{\mu} \left[ f_0 + f_0 (1 \mp f_0) \frac{p^{\mu} p^{\nu}
  \pi_{\mu \nu}}{2 T^2 (e + P)} \right]
\end{equation}
to integrate over the freezeout surface (the spacetime surface at which the matter produced in the collision cools to the freezeout temperature that we take to be 105 MeV) and obtain the momentum distribution for hadrons
with different charges.  Here,
$g$ is the hadron's spin degeneracy factor and the equilibrium distribution function is given by
\begin{equation}
  f_{0} = \frac{1}{\exp ((p \cdot V)/T) \pm 1} .
\end{equation}

With the momentum distribution for hadrons with different charge in hand, the final step in the calculation is the evaluation of the anisotropic flow coefficients as function of rapidity:
\begin{equation}
v_n (y) \equiv \frac{\int dp_T d\phi \, p_T \frac{dN}{dy p_T dp_T d\phi} \cos\left[n (\phi - \Psi_n)\right]}{\int dp_T d\phi \, p_T\frac{dN}{dy p_T dp_T d\phi}}
\label{eq5}
\end{equation}
where $\Psi_n = 0$ is the event-plane angle in the numerical simulations.
In order to define the sign of the rapidity-odd directed flow $v_1$, we choose 
the spectators at positive $x$ to fly toward 
negative $z$, as illustrated in Fig.~\ref{figschema}.
We can then compute the odd component of $v_1(y)$ according to
\begin{equation}
v_1^\mathrm{odd} = \frac{1}{2} (v_1(\Psi_+) - v_1(\Psi_-)),
\label{eq6}
\end{equation}
Experimentally, the rapidity-odd directed flow $v_1^\mathrm{odd}$ is measured \cite{Margutti:2017lup} by correlating the directed flow vector of particles of interest, ${\bf Q}^\mathrm{POI}_1 = \sum_{j=1}^{M^\mathrm{POI}} e^{i \phi_j}$, with the flow vectors from the energy deposition of spectators in the zero-degree calorimeter (ZDC), ${\bf Q}^\mathrm{ZDC}_{\pm} = \sum_j E^{\pm}_j r_j e^{i \phi_j}$.
The directed flow is defined using the scalar-product method:
\begin{equation}
v_1(\Psi_{\pm}) = \frac{1}{\langle M^\mathrm{POI} \rangle_\mathrm{ev}}\frac{\langle {\bf Q}^\mathrm{POI}_1 \cdot ( {\bf Q}^\mathrm{ZDC}_{\pm})^*\rangle_\mathrm{ev}}{\sqrt{\langle \vert {\bf Q}^\mathrm{ZDC}_{+} \cdot ({\bf Q}^\mathrm{ZDC}_-)^* \vert \rangle_\mathrm{ev}}}\ .
\label{eq15}
\end{equation}
In the definition of ${\bf Q}^\mathrm{ZDC}_{\pm}$, the index $j$ runs over all the segments in the ZDC and $E_j$ denotes the energy deposition at ${\bf x}_j = r_j e^{i \phi_j}$. 
In our notation, the flow vector angle $\Psi_+ = \pi$ in the forward ($+z$ direction) ZDC and $\Psi_- = 0$ in the backward ($-z$) direction ZDC. 
The odd component of $v_1(y)$ that we compute according to Eqs. (\ref{eq5}) and (\ref{eq6}) 
can be directly compared to $v_1^{\rm odd}$ defined from the experimental definition of 
$v_1(\Psi_\pm)$ in (\ref{eq15}).

In order to isolate the small contribution to the various flow observables that was induced by the electromagnetic fields, separating it from the much larger background hydrodynamic flow,
we compute the difference between the value of a given flow observable for positively and negatively charged hadrons:
\begin{equation}
\Delta \langle p_T \rangle \equiv \langle p_T \rangle (h^+) - \langle p_T \rangle (h^-)
\end{equation}
and
\begin{equation}
\Delta v_n \equiv v_n(h^+) - v_n(h^-),
\label{eq17}
\end{equation}
are the quantities of interest.

\section{Electromagnetic fields}
\label{sec::EM}

It is instructive to analyze the spatial distribution and the evolution of the electromagnetic fields in heavy-ion collisions. We shall do so in this Section, before turning to a discussion of the results of our calculations in the next Section.

 \begin{figure}[t!]
 \begin{center}
\includegraphics[scale=0.40]{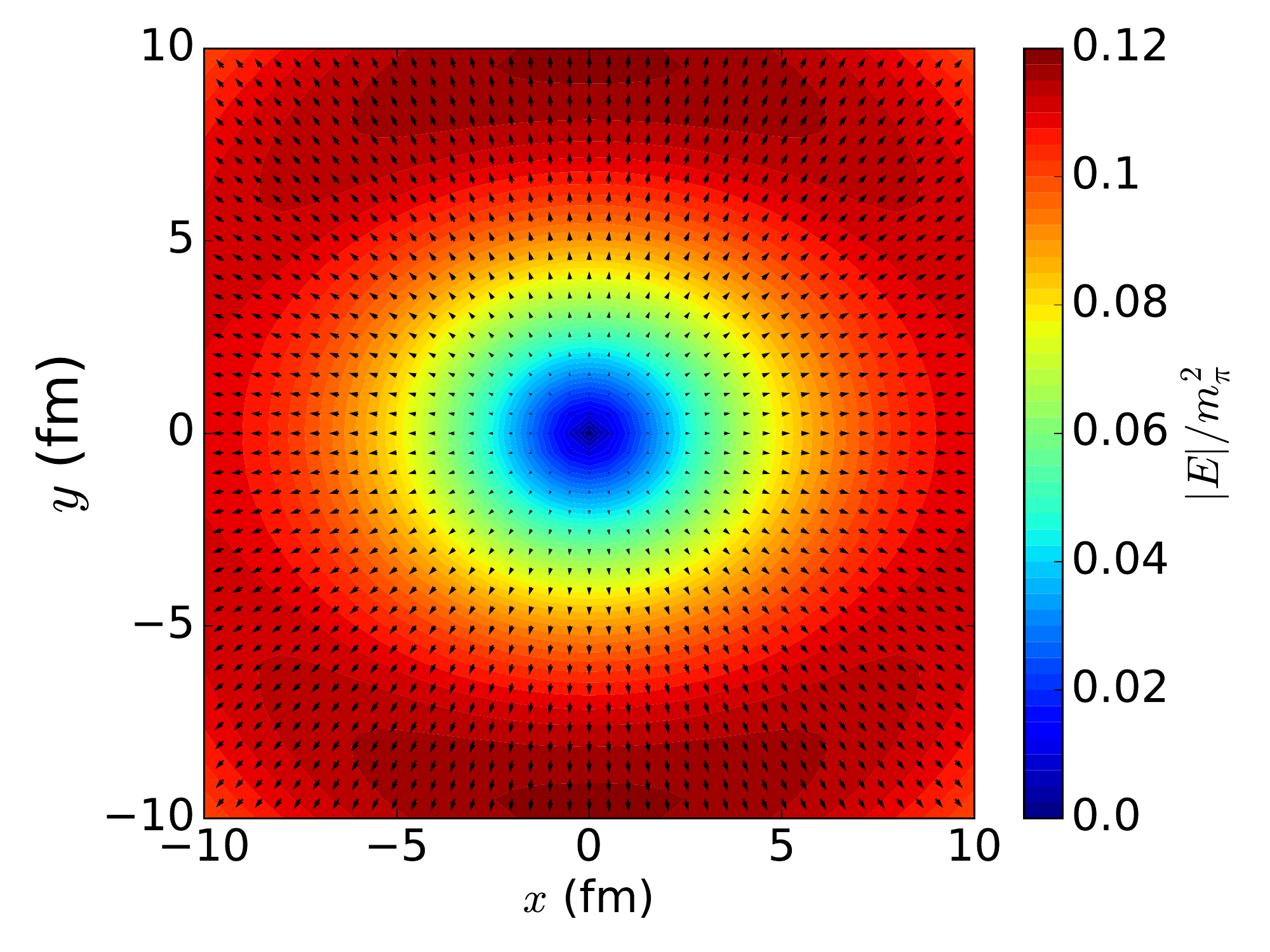}
\includegraphics[scale=0.40]{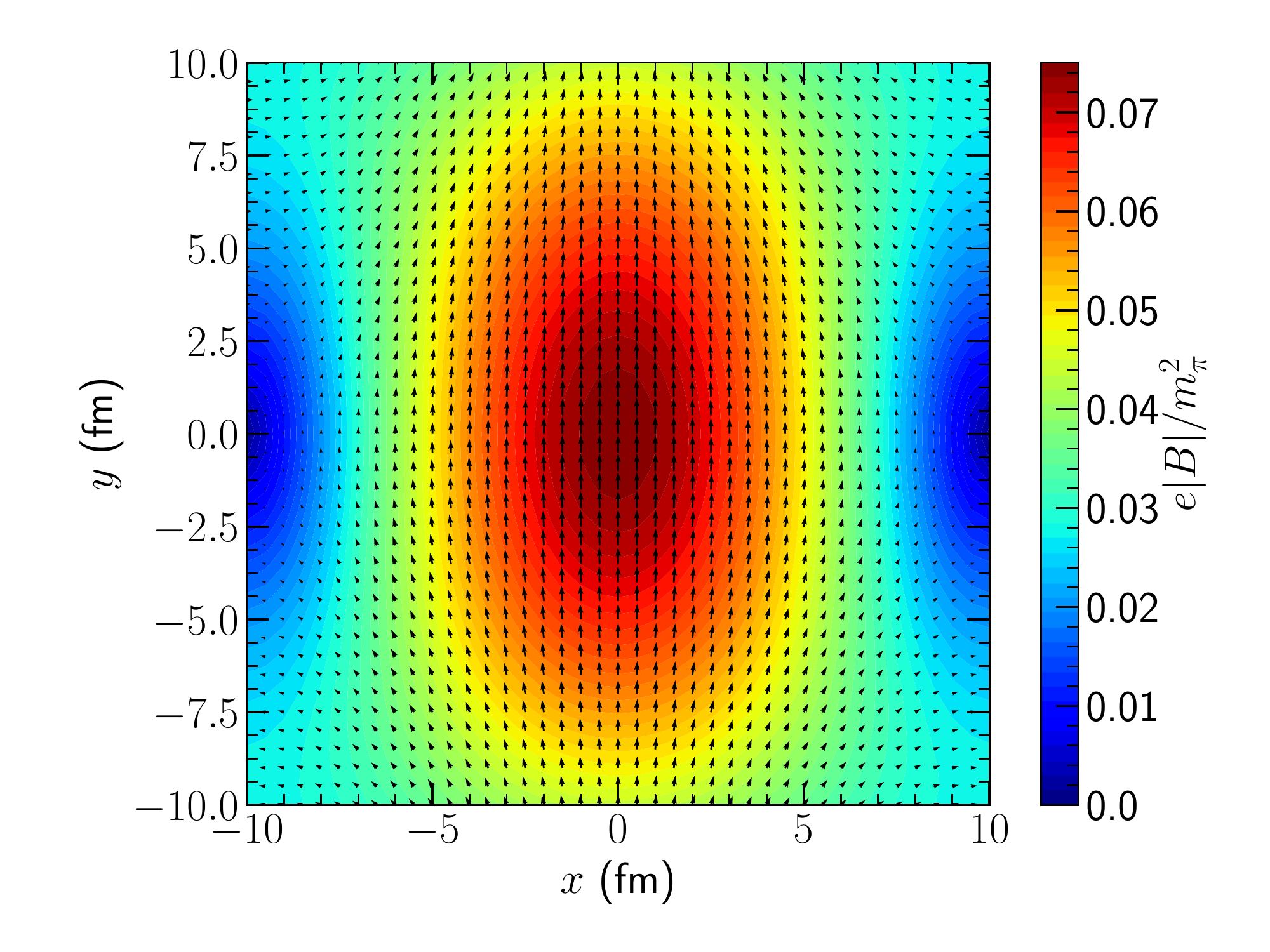}
 \end{center}
 \caption[]{The electric (left) and magnetic (right) fields in the transverse plane at $z=0$  in the lab frame 
 at a proper time $\tau=1$~fm/c after a Pb+Pb collision
 with 20-30\% centrality (corresponding to impact parameters in the range 6.24~fm~$<b<$~9.05~fm) and with a collision energy $\sqrt{s} =2.76$~ATeV.  The fields are produced by the spectator ions
 moving in the $+z$ ($-z$) direction for $x<0$ ($x>0$) as well as by the ions that participate in
 the collision. In both panels, the contribution from the spectators is larger, however. The direction of the fields are shown by the black arrows. The strength of the field is indicated both by the length of the arrows and by the color.  We see that the magnetic field is strongest at the center of the plasma, where it points in the $+y$ direction as anticipated in Fig.~\ref{figschema}. The electric field points in a generally outward direction and is strongest on the periphery of the plasma. Its magnitude is not azimuthally symmetric:
the field is on average stronger where it is pointing in the $\pm y$ directions than where it is pointing in the $\pm x$ directions.  
 }
\label{figEB}
\end{figure}

Fig.~\ref{figEB} presents our calculation of 
the magnitude and direction of the electromagnetic fields, both electric and magnetic,
in the lab frame across the $z=0$ transverse plane at a proper time
$\tau = 1$fm/c after a Pb+Pb collision with 20-30\% centrality and a collision energy of $\sqrt{s} =2.76$~ATeV. 
These electric and magnetic fields are produced by both spectator and participant ions in the two incoming nuclei. We outlined the calculation in Section \ref{sec::model}; it follows Ref.~\cite{Gursoy:2014aka}.
The spectator nucleons give the dominant contributions to the $\vec B$ field. 
The beam directions for the ions at $x>0$ ($x<0$) are chosen as $-z$ ($+z$), as in Fig.~\ref{figschema}.

The left panel in Fig.~\ref{figEB} includes three of the four different components of the electric field that we discussed in the Introduction, namely the electric field generated by Faraday's law $\vec E_F$, the Coulomb field sourced by the spectators $\vec E_C$, and the Coulomb field sourced by the net charge in the plasma $\vec E_P$. Their sum gives the total electric field in the lab frame, which is what is plotted. When we transform to the local rest frame of a moving fluid cell, namely the frame in which we calculate the electromagnetically induced drift velocity of positive and negative charges in that fluid cell, 
there is an additional component originating from the Lorentz force law, $\vec E_L$, as explained in the Introduction. The total electric field in the rest frame, which now also includes the $E_L$ component, is shown below in the left panel of Fig.~\ref{figExeta13} as a function of time.

 \begin{figure}[t!]
 \begin{center}
\includegraphics[scale=0.4]{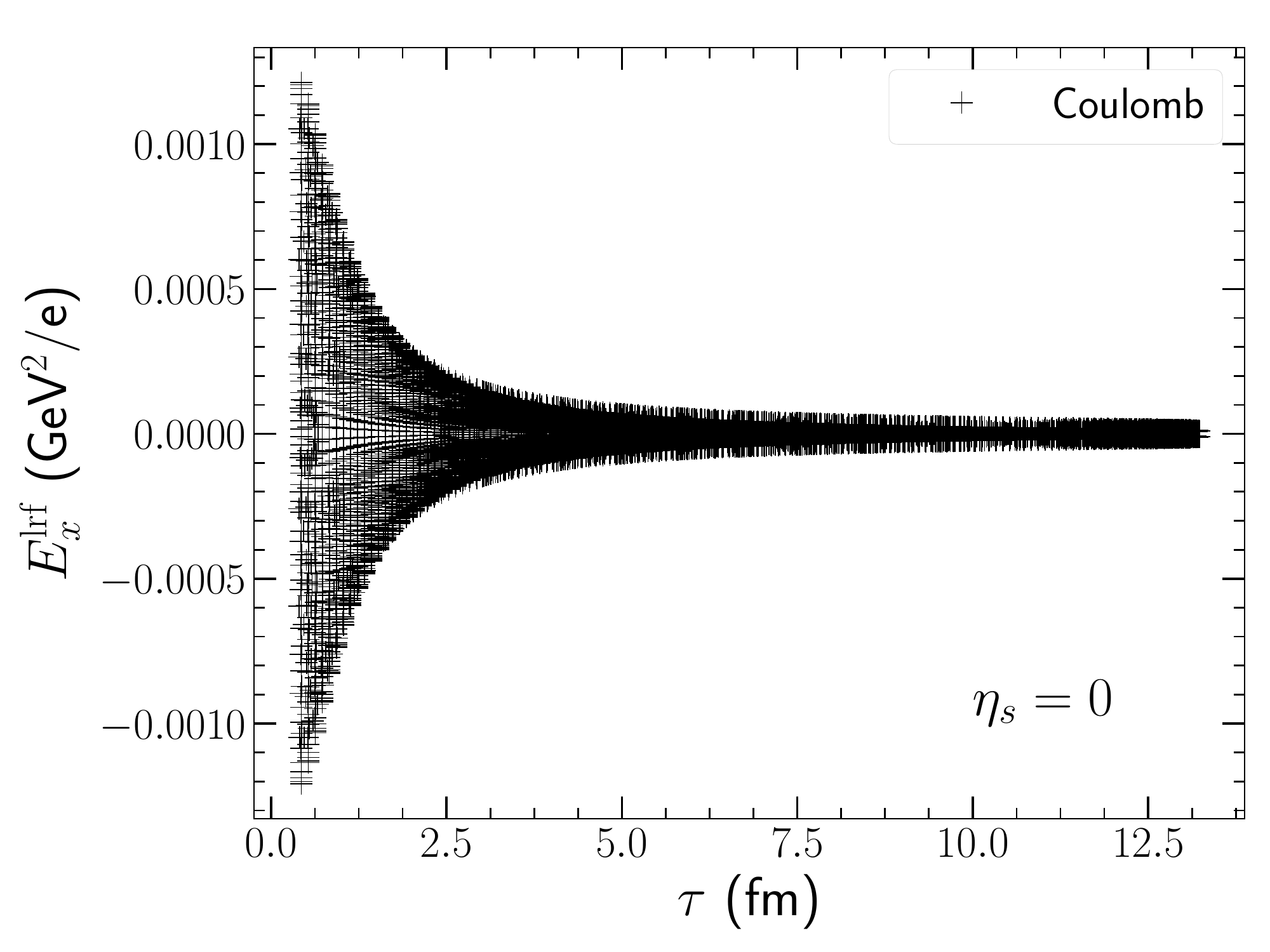}
\includegraphics[scale=0.4]{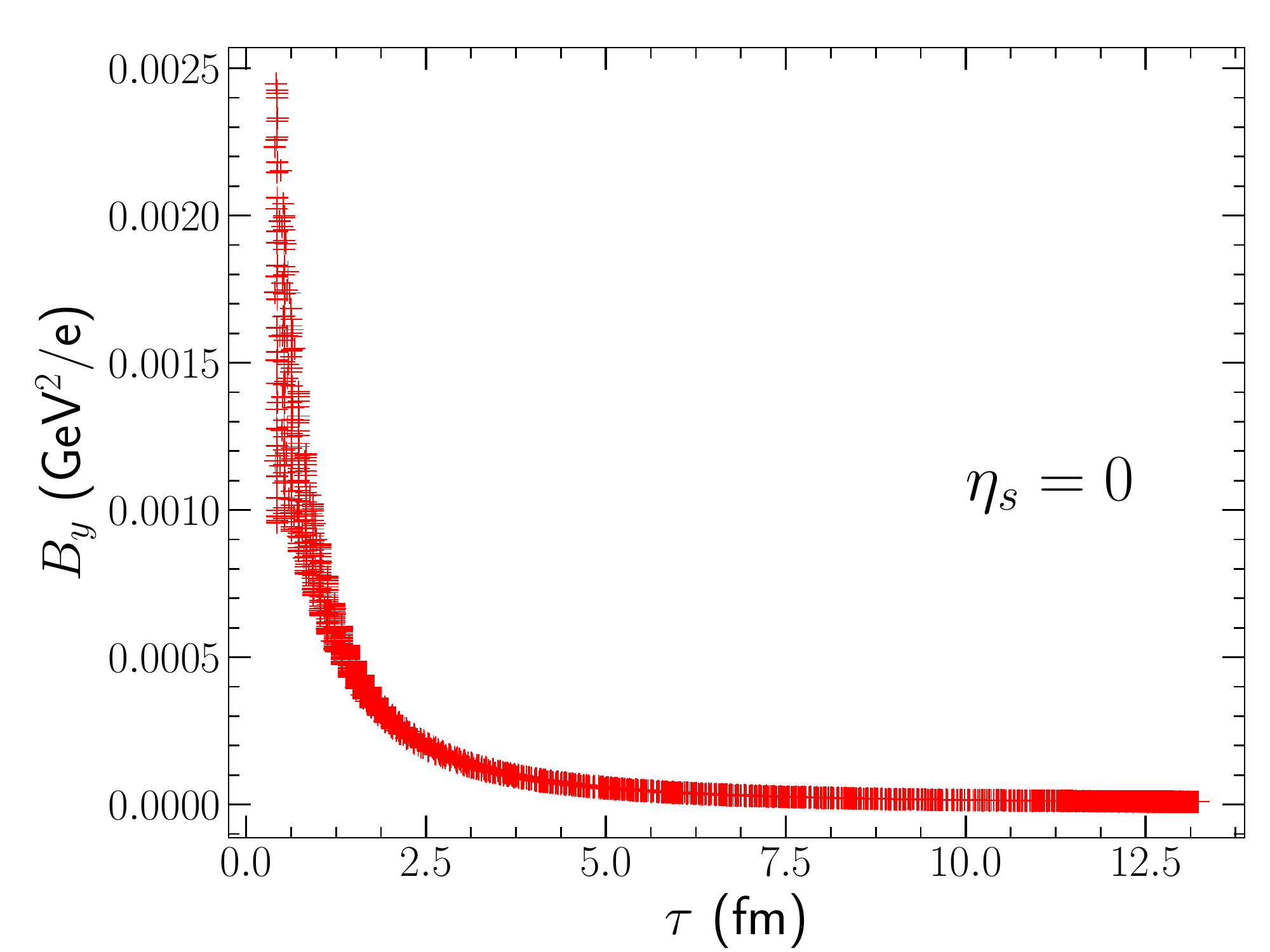}
 \end{center}
 \caption[]{Left: The $x$-component of the electric field in the local fluid rest frame at points on the freezeout surface at spacetime rapidity $\eta_s =0$, as a function of proper time. Each cross corresponds to a single fluid cell on the freezeout surface, with the vertical line of crosses at any single $\tau$ corresponding to different points on the freezeout surface at that $\tau$. Only the Coulomb electric field generated by the net charge in the plasma contributes at $\eta_s=0$, and by symmetry there for every point where $E_x^{\,\rm lrf}>0$ there is a point where $E_x^{\,\rm lrf}<0$. Right: Time dependence of the $y$-component of the magnetic field in the lab frame at $\eta_s=0$. Again, each cross corresponds to a single point on the freeze-out surface. We see that $B_y>0$ as diagrammed in Fig.~\ref{figschema} and shown in Fig.~\ref{figEB}, and here we can see how $B_y$ decreases with time.} 
\label{figEeta0}
\end{figure}
 \begin{figure}[t!]
 \begin{center}
\includegraphics[scale=0.40]{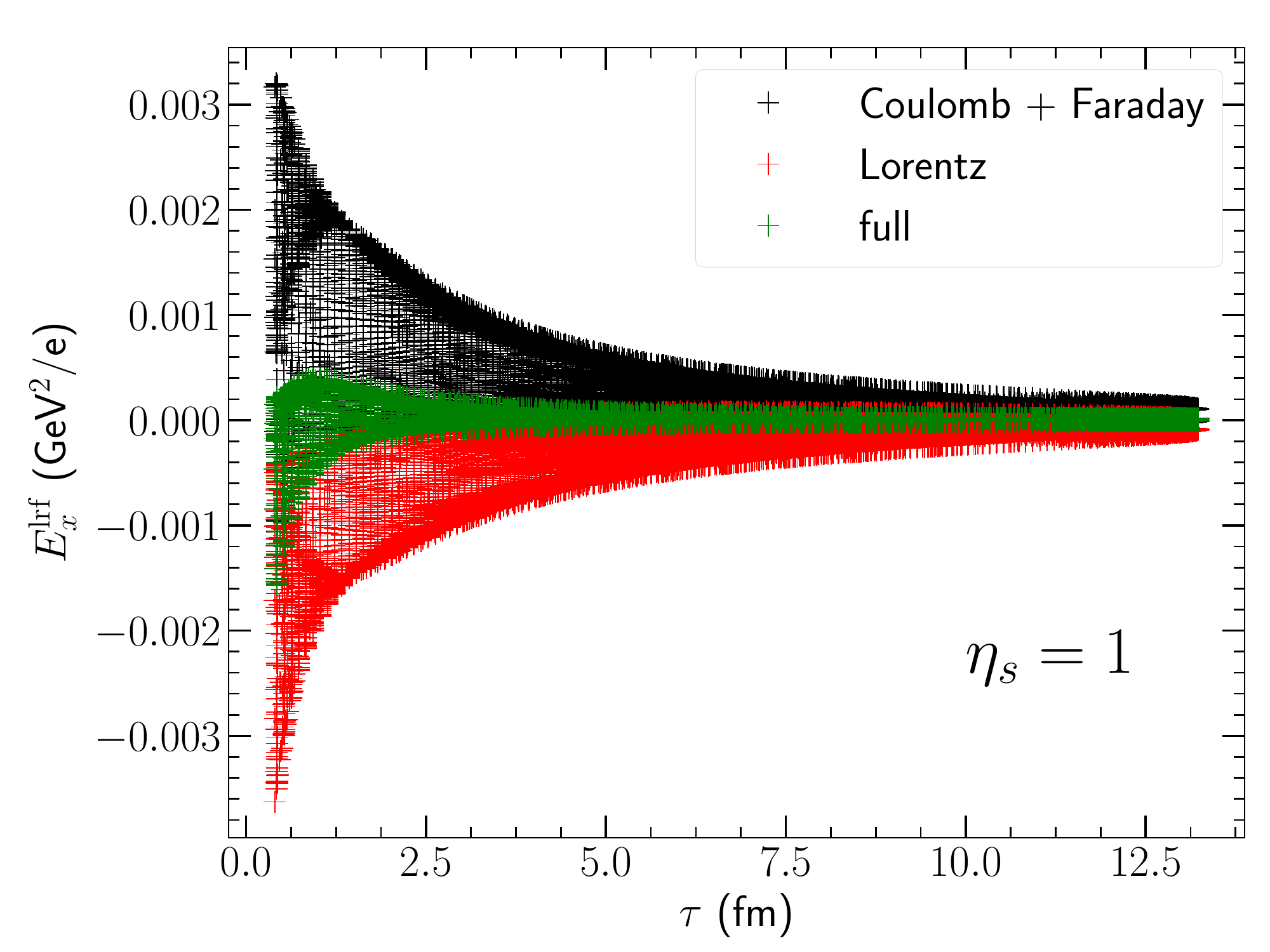}
\includegraphics[scale=0.40]{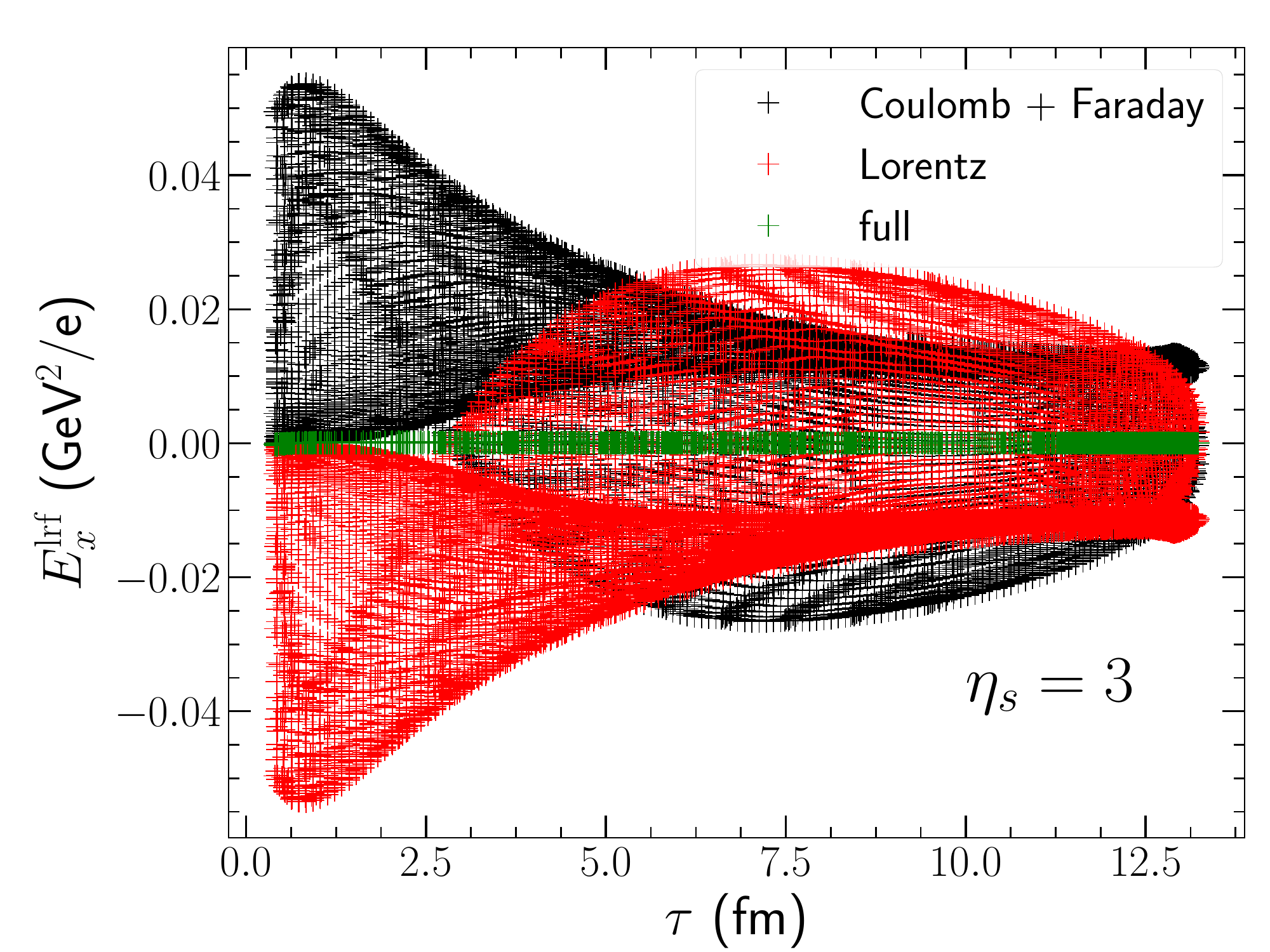}
 \end{center}
 \caption[]{Contributions to the electric field in the local rest frame of a unit cell in the fluid on the freezeout surface at a specified, non-vanishing, spacetime rapidity $\eta_s$: $\eta_s=1$ in the left panel and $\eta_s=3$ in the right panel. Each unit cell is represented in the figure by a black cross, a red cross, and a green cross.  Black crosses denote the contribution to the electric field at a given fluid cell in its local rest frame coming from the Coulomb and Faraday effects. Red crosses denote the contribution from the Lorentz force.  And, green crosses represent the total electric field at the fluid cell, namely the sum of a black cross and a red cross.
We observe that the Coulomb+Faraday and Lorentz contributions to the electric field point in opposite directions, as sketched in Fig.~\ref{figschema}, and furthermore see that the two contributions  almost cancel at large $\eta_s$, as we shall discuss in Section~\ref{sec::cancel}. We shall see
there that the Coulomb+Faraday contribution is slightly larger in magnitude than the Lorentz contribution.
 }
\label{figExeta13}
\end{figure}

The magnetic field in the right panel of Fig.~\ref{figEB} indeed decays as a function of time as shown in the right panel of  Fig.~\ref{figEeta0}. Via Faraday's law this induces a current in the same direction as the current pushed by the Coulomb electric field coming from the spectators, and it opposes the current caused by the Lorentz force on fluid elements moving in the longitudinal direction, 
as sketched in Fig.~\ref{figschema} and seen in Fig.~\ref{figExeta13}.

When solving the force-balance equation, Eq.~(\ref{eq9}), we find that the drift velocity is 
mainly determined by the electric field in the local local fluid rest frame. 
To understand how the Coulomb, Lorentz and Faraday effects contribute to the drift velocity 
on the freeze-out surface it is instructive to study how the different effects contribute to 
the electric field in the local fluid rest frame.  We do so at $\eta_s=0$ in the left panel of Fig.~\ref{figEeta0}.  At $\eta_s=0$, only the Coulomb effect contributes.  This means that when in Section~\ref{sec::results} we compute the charge-odd contribution to the even flow harmonics at $\eta_s=0$ this will provide an estimate of the magnitude of the Coulomb contribution to the flow coefficients.
In Fig.~\ref{figExeta13} we look at the different contributions to the electric field in the local fluid rest frame at $\eta_s=1$ and $\eta_s=3$.
We see that the Coulomb + Faraday and Lorentz effects point in opposite directions, and almost cancel at large spacetime rapidity. We discuss the origin and consequences of this cancellation in Section \ref{sec::cancel} below. 

\section{Results} 
\label{sec::results}

In this Section we present our results for the charge-dependent contributions to the 
anisotropic flow  induced by the electromagnetic effects introduced in Section~\ref{sec::intro}.
As we have described in Section~\ref{sec::model}, to obtain the anisotropic flow coefficients we input the electromagnetic fields in the local rest frame of the fluid, calculated in Section~\ref{sec::EM}, into the force-balance equation (\ref{eq9}) which then yields the electromagnetically induced component of the velocity field of the fluid. This velocity field is then input into the Cooper-Frye freezout procedure~\cite{Cooper:1974mv} to obtain the distribution of particles in the final state and, in particular, the anisotropic flow coefficients~\cite{Gursoy:2014aka}. 

\begin{figure*}[t!]
  \begin{tabular}{cc}
  \includegraphics[width=0.48\linewidth]{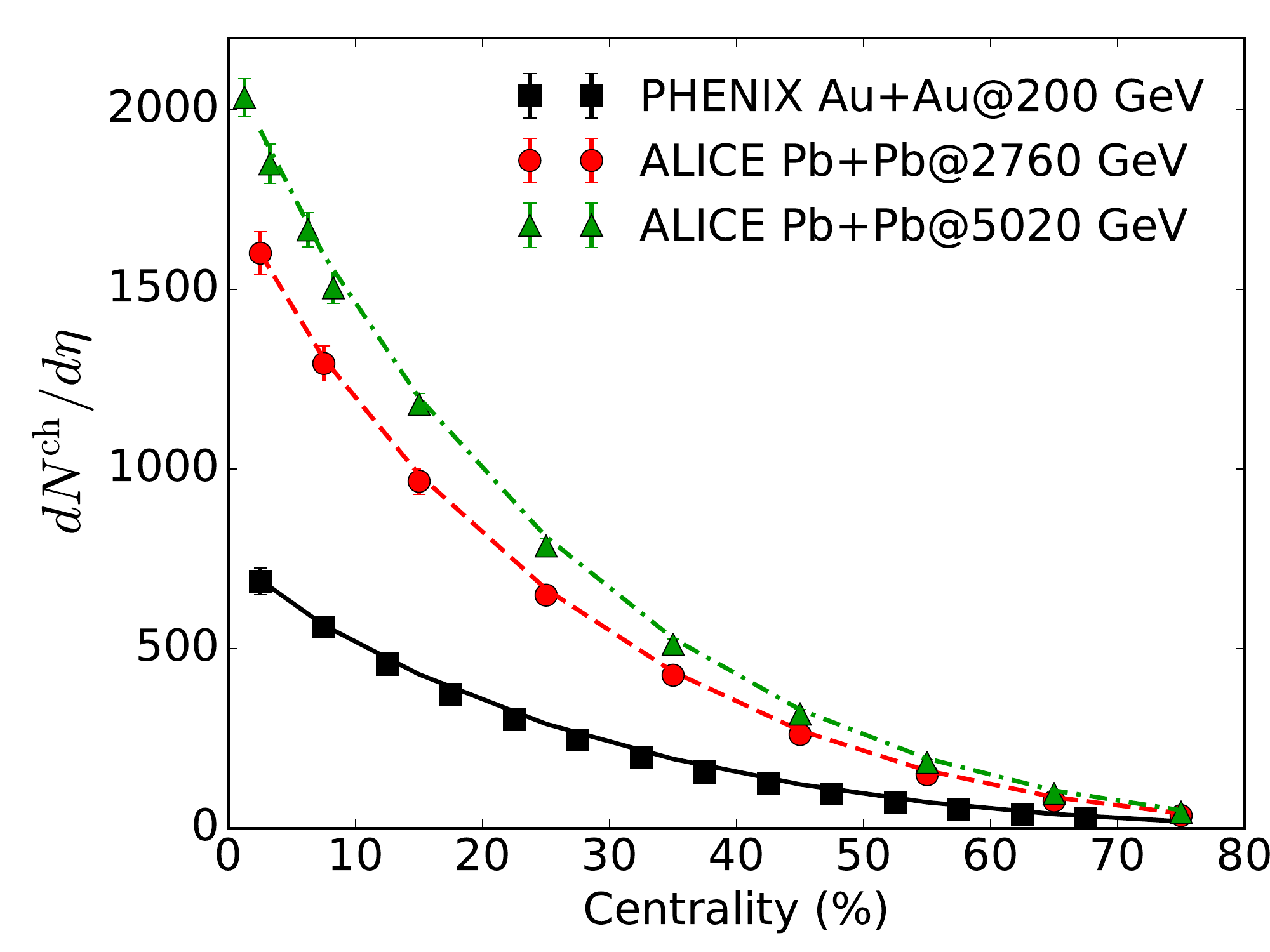} &
   \includegraphics[width=0.48\linewidth]{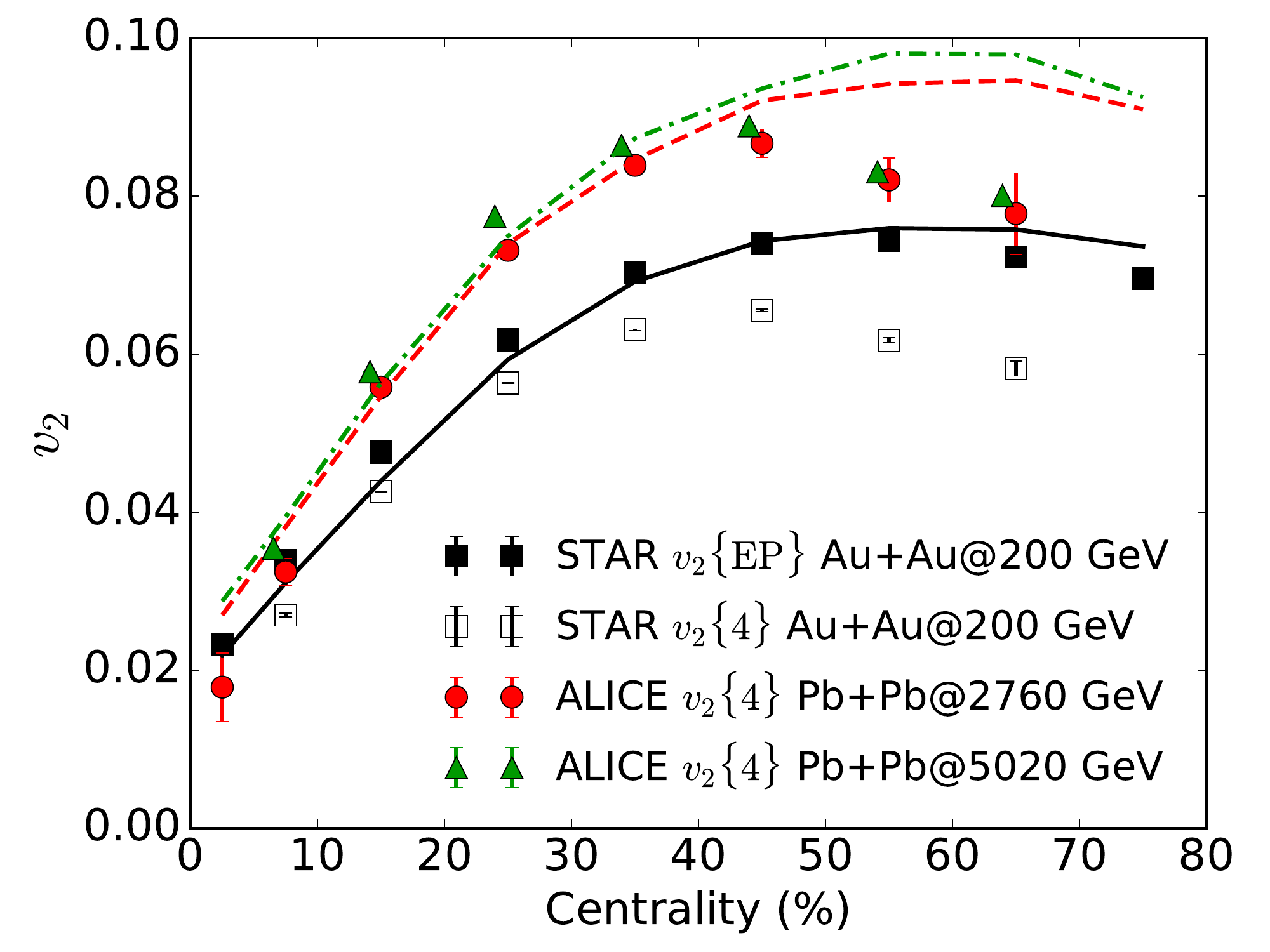}  
  \end{tabular}
  \caption{To get a sense of how well the solution to relativistic viscous hydrodynamics upon which we build our calculation of electromagnetic fields and currents describes heavy ion collisions, we compare our results for charged hadron multiplicities (left) and elliptic flow coefficients (right) to  experimental measurements at the top RHIC and LHC energies from Refs. \cite{Adler:2004zn,Aamodt:2010cz,Adam:2015ptt} and Refs. \cite{Adams:2004bi,Aamodt:2010pa,Adam:2016izf}, respectively. }
  \label{fig1}
\end{figure*}

To provide a realistic dynamical background on top of which to compute the electromagnetic fields and consequent currents, we have calibrated the solutions to relativistic viscous hydrodynamics that we use by comparing them to experimental measurements of hadronic observables. To give a sense of the agreement that we have obtained, in Fig.~\ref{fig1} we show our results for
the centrality dependences of charged hadron multiplicity and elliptic flow coefficients are shown  for heavy-ion collisions at three collision energies as well as data from STAR, PHENIX and ALICE Collaborations \cite{Adler:2004zn,Aamodt:2010cz,Adam:2015ptt,Adams:2004bi,Aamodt:2010pa,Adam:2016izf}. 
Since we do not have event-by-event fluctuations in our calculations, we compare our results for the elliptic 
flow coefficent $v_2$ to experimental measurements of $v_2$ from the 4-particle cumulant, 
$v_2\{4\}$~\cite{Qiu:2011iv}. With the choice of the specific shear viscosity $\eta/s(T)$ that we have made in Eq.~(\ref{eq1}), our model provides a reasonable agreement with charged hadron $v_2\{4\}$ for heavy ion collisions with centralities up to the 40-50\% bin.  

To isolate the effect of electromagnetic fields on charged hadron flow observables, 
we study the difference between the $v_n$ of positively charged particles and the $v_n$ of negatively charged particles as defined in Eq. (\ref{eq17}). We also study the difference between the mean transverse momentum $\langle p_T \rangle$ of positively charged hadrons and that of negatively charged hadrons. This provides us with information about the modification in the hydrodynamic radial flow induced by the electromagnetic fields. The difference between the charge-dependent flow of light pions and heavy protons is also compared. Hadrons with different masses have different sensitivities to the underlying hydrodynamic flow and 
to the electromagnetic fields.

We should distinguish the charge-odd contributions to the 
odd flow moments, $\Delta v_1$, $\Delta v_3$, $\ldots$, from 
the charge-dependent contributions to the
even ones, $\Delta v_2$, $\Delta v_4$, $\ldots$, as they have qualitatively different origins.
The charge-odd contributions to the odd flow coefficients induced by electromagnetic fields, $\Delta v_{2n-1}$, 
are rapidity-odd: $\Delta v_{2n-1}(\eta_s) = - \Delta v_{2n-1}(-\eta_s)$. 
This can easily be understood by inspecting Fig.~\ref{figschema}, where we describe different effects that contribute to the total the electric field in the plasma. This can also be proven analytically by studying the transformation property of $\Delta v_n$ under $\eta \to -\eta$. 
As we have seen in Section~\ref{sec::intro}, there are three basic effects that contribute. First, there is the electric field produced directly by the positively charged spectator ions. They generate electric fields in opposite directions in the $z>0$ and $z<0$ regions. 
We call this the {\em Coulomb electric field} $\vec{E}_C$, as the resulting electric current in the plasma is a direct result of the Coulomb force between the spectators and charges in the plasma. Then there are the two separate magnetically induced electric fields, as discussed 
in Ref.~\cite{Gursoy:2014aka}. The {\em Faraday electric field} $\vec{E}_F$ results from the rapidly decreasing magnitude of the magnetic field perpendicular to the reaction plane, see Fig.~\ref{figschema}, as a consequence of Faraday's law. 
Note that $\vec{E}_F$ and $\vec{E}_C$ point in the same directions. 
Finally, there is another magnetically induced electric field, the {\em Lorentz electric field} $\vec{E}_L$ that can be described in the lab frame as the Lorentz force on charges that are moving because of the longitudinal expansion of the plasma and that are in a magnetic field.  Upon transforming to the local fluid rest frame, the lab-frame magnetic field becomes an electric field that we denote $\vec{E}_L$.\footnote{This electric field was called the Hall electric field in Ref.~\cite{Gursoy:2014aka}.} As shown in Fig.~\ref{figschema}, $\vec{E}_L$ points  in the opposite direction to $\vec{E}_F$ and $\vec{E}_C$.

\begin{figure*}[t!]
  \includegraphics[width=0.8\linewidth]{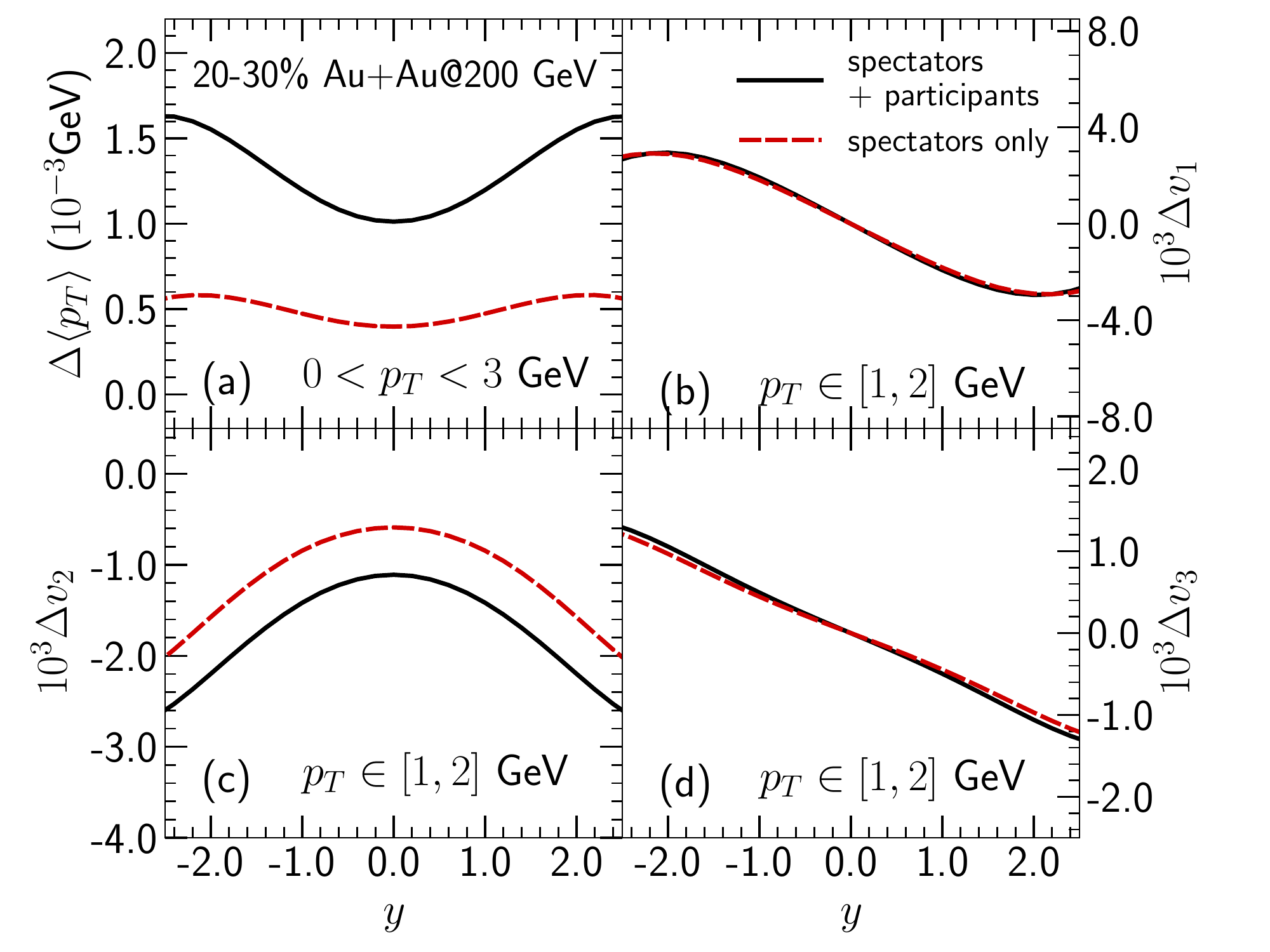}   
  \caption{The solid black curves display the principal results of our calculations for 20-30\% centrality Au+Au collisions at 200 GeV, as at RHIC.  We show the contribution to the mean-$p_T$ of charged pions and the first three $v_n$ coefficients induced by the electromagnetic fields that we have calculated, isolating the electromagnetically induced effects by taking the difference between the calculated value of each observable for $\pi^+$ and $\pi^-$ mesons, namely the charge-odd or charge-dependent contributions that we denote $\Delta \langle p_T \rangle$ and $\Delta v_n$.  We see rapidity-odd contributions $\Delta v_1$ and $\Delta v_3$ 
and rapidity-even contributions $\Delta \langle p_T \rangle$ and $\Delta v_2$.
 The red dashed curves show the results we obtain when we calculate the same observables in the presence of the electromagnetic fields produced by the spectators only.
We see that the dominant contribution to the odd $v_n$'s is generated by these spectator-induced fields, whereas the even $v_n$'s also receive a significant contribution from the Coulomb force exerted on charges in the plasma by other charges in the plasma, originating from the participant nucleons.
}
  \label{fig:participant_effect}
\end{figure*}

On the other hand, the charge-dependent contributions to the even order anisotropic flow coefficients $v_{2n}$ are even under $\eta_s\to-\eta_s$. 
Obviously this cannot arise
from the rapidity-odd electric fields described above. 
Instead, we find that although the electromagnetic contribution to the $v_{2n}$ 
receives some contribution from components of the electric fields above that are rapidity-even and that are odd under $x \rightarrow -x$, it also receives an important contribution from the Coulomb force between the net positive electric charge in the plasma.
This arises as a result of the Coulomb force exerted on the charges in the plasma by each other ---
as opposed to the Coulomb force exerted on charges in the plasma by the spectator ions.
This electric field is non-trivial even at $z=0$ as shown in Fig.~\ref{figEB} (left). 
We call this field the {\em plasma electric field} and denote it by $\vec{E}_P$. 
This contributes to the net $\Delta v_2$ and it is clear from the geometry that it makes no contribution to the odd flow harmonics.

In Fig.~\ref{fig:participant_effect}, we begin the presentation of our principal results. This figure 
shows $\Delta v_n$, the charge-odd contribution to the anisotropic  flow harmonics induced by electromagnetic fields,
for pions in 20-30\% Au+Au collisions at 200 GeV. 
It also shows the difference in the mean-$p_T$ of particles with positive and negative charge, which shows how the electromagnetic fields modify the hydrodynamic radial flow. 
The radial outward pointing electric fields in Fig.~\ref{figEB} increase the radial flow for positively charged hadrons while reducing the flow for negative particles. We see that the effect is even in rapidity.
Fig.~\ref{fig:participant_effect} shows that these fields also make a charge-odd, rapidity-even
contribution to $v_2$.

We compare the red dashed curves, arising from electromagnetic effects by spectators only, with the solid black curves that show the full calculation including the participants. Noting that the lines are significantly different it follows that the Coulomb force exerted on charges in the plasma by charges in the plasma makes a large contribution to $\Delta \langle p_T \rangle$ and $\Delta v_2$.
The induced $\Delta \langle p_T \rangle$ is larger at forward and backward rapidities, because the electric fields from the spectators and from the charge density in the plasma deposited according to the distribution (\ref{ParticipantRapidityDistribution})  are both stronger there.  

The electromagnetically induced elliptic flow
$\Delta v_2$ originates from the Coulomb electric field in the transverse plane, depicted  in Fig.~\ref{figEB}. We see there that the Coulomb field is stronger along the $y$-direction than in the $x$-direction.  This reduces the elliptic flow $v_2$ for positively charged hadrons and increases it for negatively charged hadrons. Hence, $\Delta v_2$ is negative.  

Note that  $\Delta \langle p_T \rangle$ and $\Delta v_2$ are much smaller than $\langle p_T \rangle$ and $v_2$; in the calculation of Fig.~\ref{fig:participant_effect}, $\langle p_T \rangle \approx 0.47$~GeV and $v_2\approx 0.048$ for both the $\pi^+$ and $\pi^-$. The differences between these observables for $\pi^+$ and $\pi^-$ that we plot are much smaller, with $\Delta \langle p_T \rangle$ smaller than $\langle p_T \rangle$ 
by a factor of $\mathcal{O}(10^{-3})$ 
and $\Delta v_2$ smaller than $v_2$
by a factor of $\mathcal{O}(10^{-2})$ 
in Au+Au collisions at 200 GeV. 
This reflects, and is consistent with, the fact that the drift velocity induced by the 
electromagnetic fields is a small perturbation compared to the overall hydrodynamic flow on the freeze-out surface.

The electromagnetically induced contributions to the odd flow harmonics $\Delta v_1$ and $\Delta v_3$ are odd in rapidity.  In our calculation, which neglects fluctuations, $v_1$ and $v_3$ both vanish in the absence of electromagnetic effects.
We see from Fig.~\ref{fig:participant_effect} that the magnitudes of $\Delta v_1$ and $\Delta v_3$ are controlled by the electromagnetic fields due to the spectators, namely $\vec E_F$, $\vec E_C$ 
and $\vec E_L$.   By comparing the sign of the rapidity-odd $\Delta v_1$ that we have calculated in Fig.~\ref{fig:participant_effect} to the illustration in Fig.~\ref{figschema}, 
we see that 
the rapidity-odd electric current flows in the direction of $\vec E_F$ and $\vec E_C$, opposite to the direction of $\vec E_L$, meaning that $|\vec E_F+\vec E_C|$ is greater than $|\vec E_L|$.  Our results for $\Delta v_1$ are qualitatively similar to those found in  Ref.~\cite{Gursoy:2014aka}, although they differ quantitatively because of the differences between our realistic hydrodynamic background and the simplified hydrodynamic solution used in Ref.~\cite{Gursoy:2014aka}.
Here, we find a nonzero $\Delta v_3$ in addition, also odd in rapidity, and with the same sign 
as $\Delta v_1$  and a similar magnitude.  This is natural since $\Delta v_3$ receives a contribution from the mode coupling between the electromagnetically induced $\Delta v_1$ and the background elliptic flow $v_2$.

\begin{figure*}[t!]
  \includegraphics[width=0.8\linewidth]{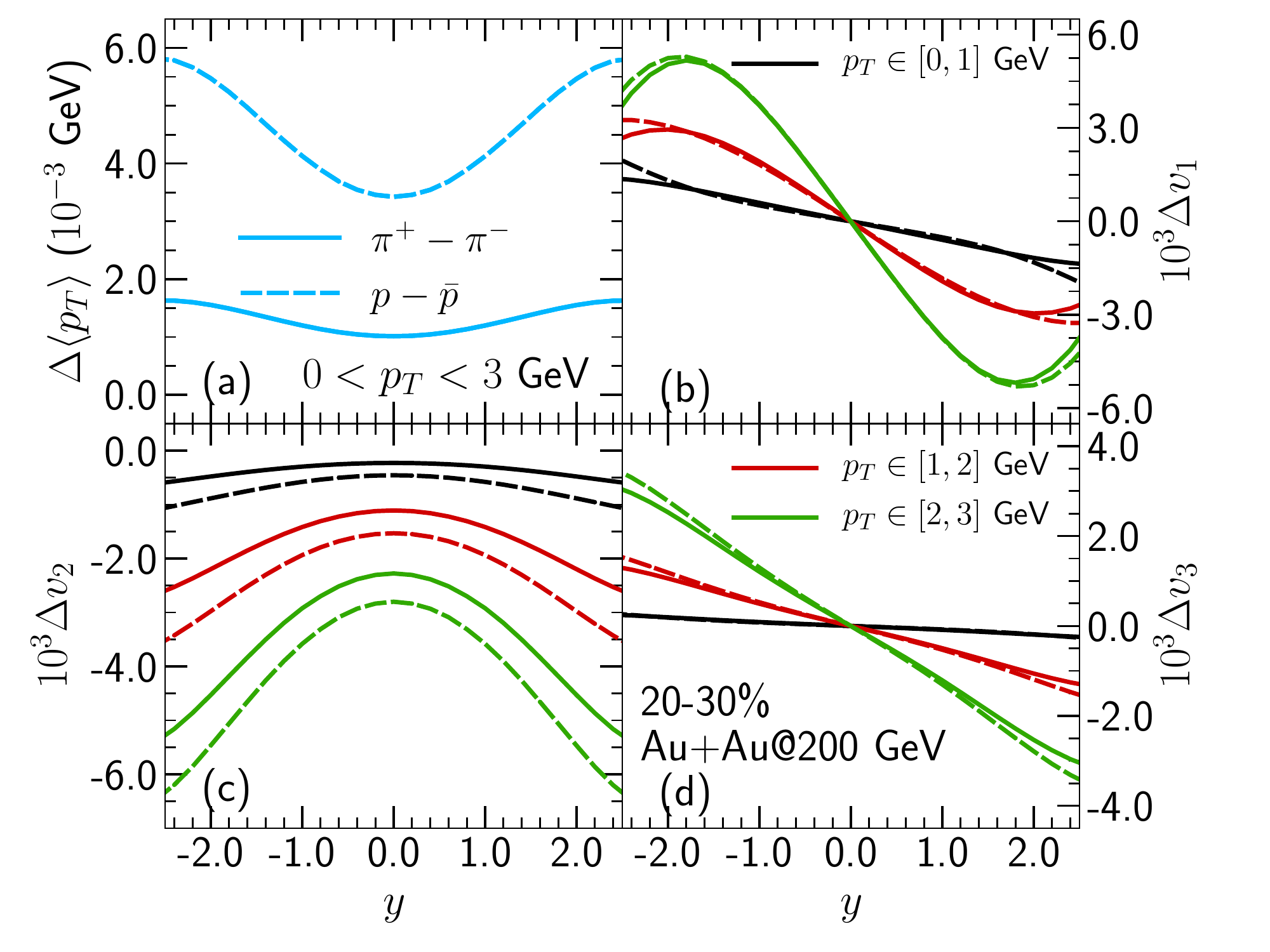}   
  \caption{The electromagnetically induced difference between the mean $p_T$ and $v_n$ coefficients of $\pi^+$ and $\pi^-$ mesons (solid lines) and between protons and antiprotons (dashed lines) as a function of particle rapidity for 20-30\% Au+Au collisions at 200 GeV. Three different $p_T$ integration ranges are shown for each of the $\Delta v_n$ as a function of particle rapidity.}
  \label{fig2}
\end{figure*}

In Fig.~\ref{fig2} we see that the heavier protons have a larger electromagnetically induced 
shift in their mean $p_T$ compared to that for the lighter pions. Because a proton has a larger mass than a pion, its velocity is slower than that of a pion with the 
same transverse momentum, $p_T$. 
Thus, when we compare pions and protons with the same $p_T$, 
the hydrodynamic radial flow generates a stronger blue shift effect for the less relativistic proton spectra, which is to say that the proton spectra are more sensitive to the hydrodynamic radial flow~\cite{Heinz:2004qz}. Similarly, when the electromagnetic fields that we compute 
induce a small difference between the radial flow velocity of positively charged particles relative to that of negatively charged particles,
the resulting difference between the mean $p_T$ of protons and antiprotons is greater than the difference between the mean $p_T$ of positive and negative pions. Turning to the $\Delta v_n$'s, 
we see in Fig.~\ref{fig2} that the difference between the electromagnetically induced $\Delta v_n$'s for protons and those for pions are much smaller in magnitude. We shall also see below that these differences are modified somewhat by contributions from pions and protons produced after freezeout by the decay of resonances.  For both these reasons, these differences cannot be interpreted via a simple blue shift argument.
Fig.~\ref{fig2} also shows the charge-odd electromagnetically induced flow coefficients $\Delta v_n$
computed from charged pions and protons+antiprotons in three different $p_T$ ranges. The $\Delta v_1$, $\Delta v_2$ and $\Delta v_3$ all increase as the $p_T$ range increases, in much the same 
way that the background $v_2$ does. In the case of 
$\Delta v_1$, this agrees with what was found in Ref.~\cite{Gursoy:2014aka}. 

\begin{figure*}[t!]
  \includegraphics[width=0.8\linewidth]{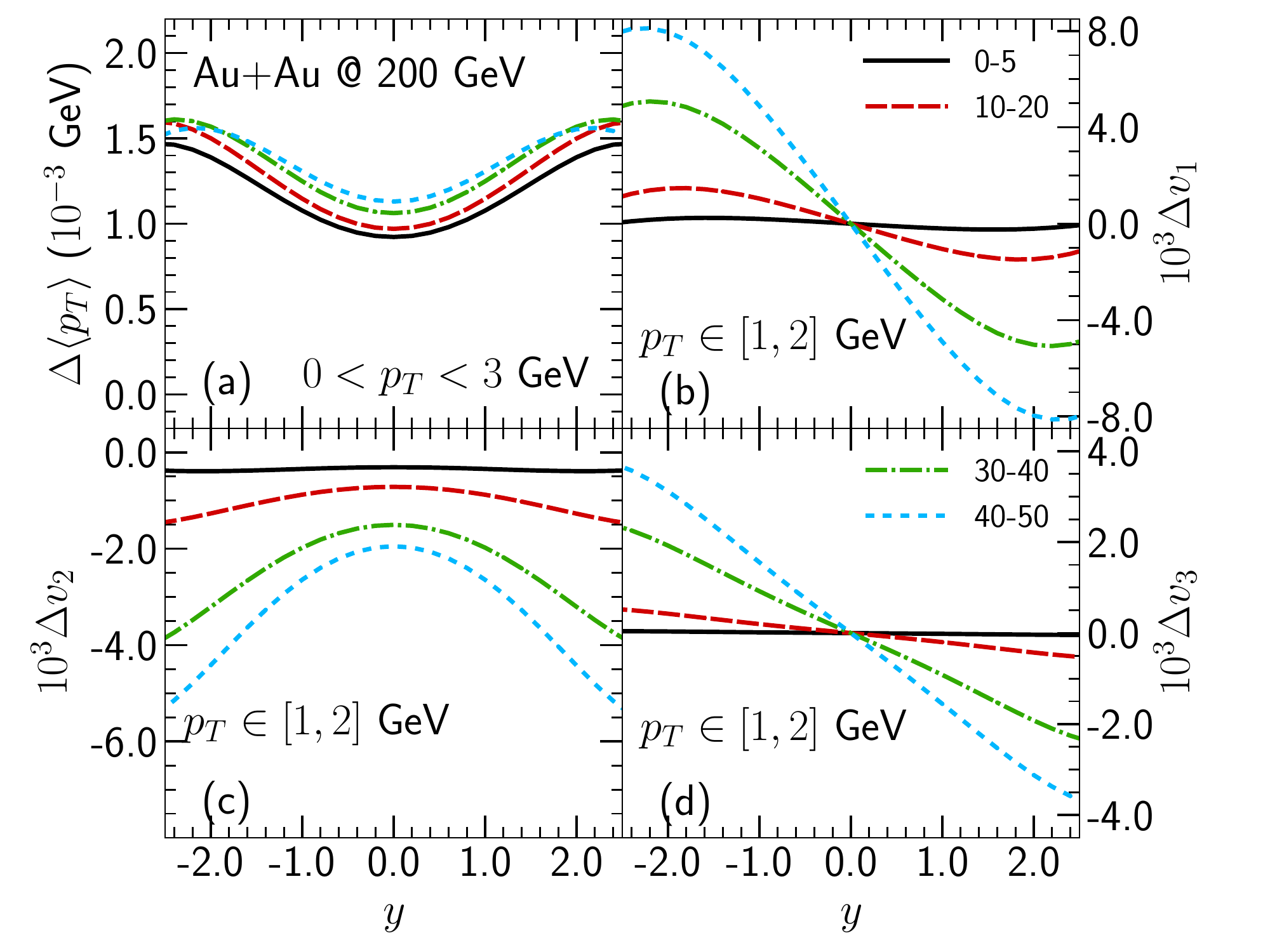}   
  \caption{The centrality dependence of the electromagnetically induced flow difference in $\pi^+$ vs $\pi^-$ as a function of particle rapidity in Au+Au collisions at 200 GeV.}
  \label{fig3}
\end{figure*}

In Fig.~\ref{fig3} we study the centrality dependence of the  electromagnetically induced  flow in Au+Au collisions at 200 GeV. 
The difference between the flow of positive and negative pions, both the radial flow and the flow anisotropy coefficients, increases as one goes from central toward peripheral heavy ion collisions. 
However, the increase in $\Delta \langle p_T \rangle$ and $\Delta v_2$ is smaller than 
the increase in the odd
$\Delta v_n$'s.  This further confirms that the odd $\Delta v_n$'s are induced by the electromagnetic fields produced by the spectator nucleons only -- since the more peripheral a collision is the more spectators there are.  

Compared to any of the anisotropic flow coefficients $\Delta v_n$, the $\Delta \langle p_T \rangle$ shows the least centrality dependence because, as we saw in Fig.~\ref{fig:participant_effect},
$\Delta \langle p_T \rangle$ originates largely from the Coulomb field of the plasma, coming from the charge of the participants, with only a small contribution from the spectators.
The increase of $\Delta v_2$ with centrality is intermediate in magnitude, since it originates both
from the participants and from the spectators, as seen in Fig.~\ref{fig:participant_effect}.
Another origin for the increase in electromagnetically induced effects in more peripheral collisions
is that the typical lifetime of the fireball in these collisions is shorter compared to that in central collisions. This gives less time for the electromagnetic fields to decay by the time of peak 
particle production in more peripheral collisions. 
In the case of  $\Delta \langle p_T \rangle$, which is dominantly controlled by the plasma Coulomb field which is less in more peripheral collisions where there is less plasma, this effect partially cancels the effect of the reduction in the fireball lifetime, and
results in $\Delta \langle p_T \rangle$ being almost centrality independent.

\begin{figure}[t!]
  \includegraphics[width=0.46\linewidth]{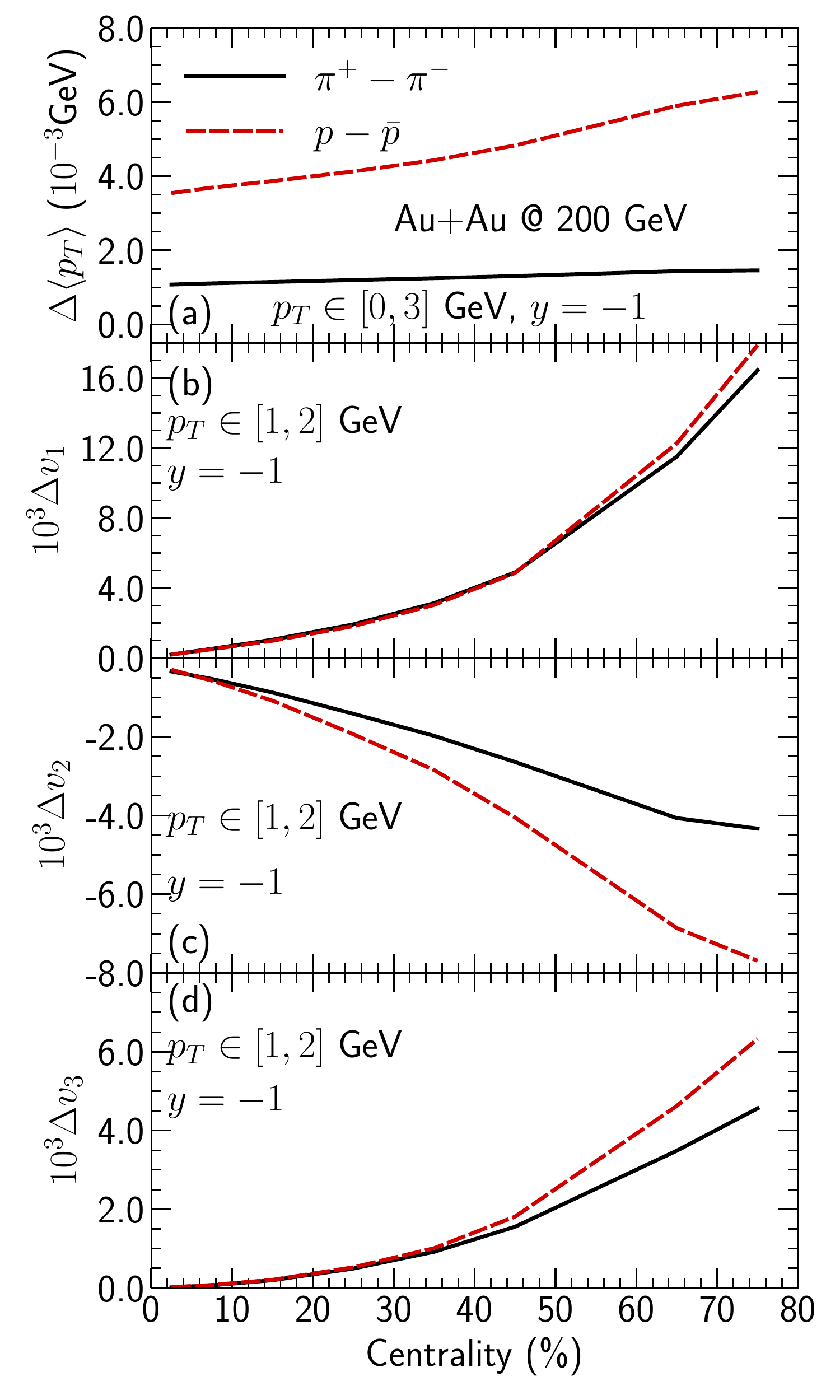}   
  \caption{The centrality dependence of the electromagnetically induced differences in the radial flow and anisotropic flow coefficients for positively and negatively charged hadrons, here at a fixed rapidity $y=-1$.}
  \label{fig4}
\end{figure}
%
Fig.~\ref{fig4} further shows the centrality dependence of the  electromagnetically induced difference between flow observables for positive and negative particles at a fixed rapidity. 
We observe that  $\Delta \langle p_T \rangle$ 
does not vanish in central collisions. This further confirms that it is largely driven by
the Coulomb field created by a net positive charge density in the plasma itself, as this Coulomb field is present in collisions with zero impact parameter whereas all spectator-induced effects vanish when there are no spectators.
This charge density creates an outward electric field that generates an outward flux of positive charge in the plasma and leads to a non-vanishing charge-identified radial flow.

\begin{figure*}[t!]
  \includegraphics[width=0.8\linewidth]{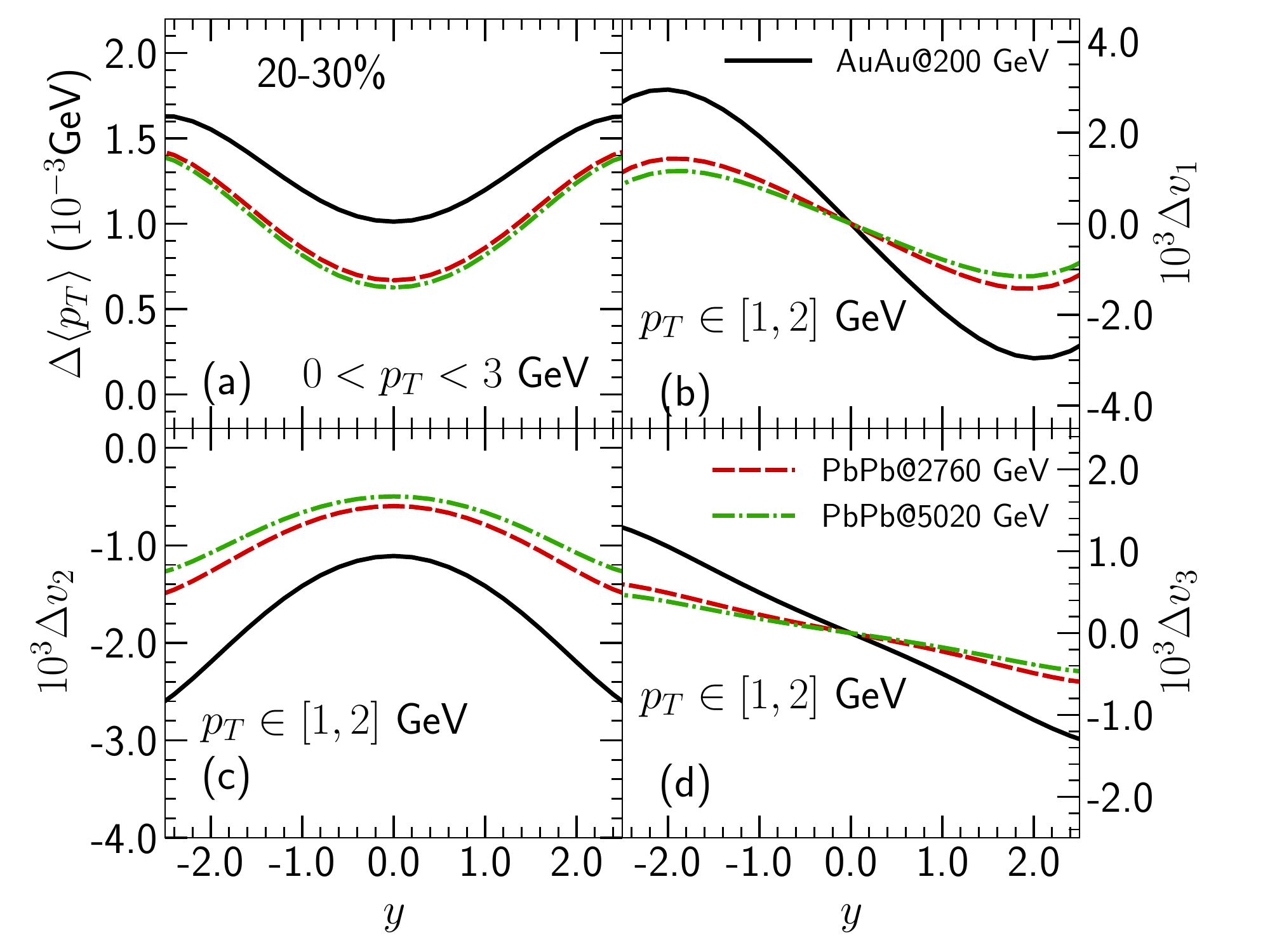}   
  \caption{The collision energy dependence of the electromagnetically induced charge-odd contributions to flow observables. The difference of particle mean $p_T$ and $v_n$ between $\pi^+$ and $\pi^-$ are plotted as a function of particle rapidity for collisions at the top RHIC energy of 200 GeV and at two LHC collision energies.}
  \label{fig5}
\end{figure*}

In Fig.~\ref{fig5}, we study the collision energy dependence of the effects of electromagnetic fields on charged hadron flow. 
The electromagnetically induced effects on the differences between flow observables for positive and negative particles are larger at the top RHIC energy than at LHC energies.
This can be understood as arising from the fact that because the spectators pass by more quickly in higher energy collisions the spectator-induced electromagnetic fields decrease more rapidly with time
in LHC collisions than in RHIC collisions.  Furthermore, in higher energy collisions at the LHC the fireball lives longer, further reducing the magnitude of the electromagnetic fields on the freeze-out surface.
The results illustrated in Fig.~\ref{fig5} motivate repeating our analysis for the lower energy collisions being done in the RHIC Beam Energy Scan, although
doing so will require more sophisticated  underlying hydrodynamic calculations and we also note that
in such collisions there are other
physical effects that contribute significantly to $\Delta \langle p_T \rangle$ and $\Delta v_2$~\cite{Xu:2012gf,Steinheimer:2012bn,Song:2012cd,Adamczyk:2013gv,Xu:2013sta,Adamczyk:2015fum,Xu:2016ihu}, in the case of $\Delta v_2$ for protons making a contribution with opposite sign to the one that we have calculated. For both these reasons, we leave such investigations to future work.

\begin{figure*}[t!]
  \includegraphics[width=0.8\linewidth]{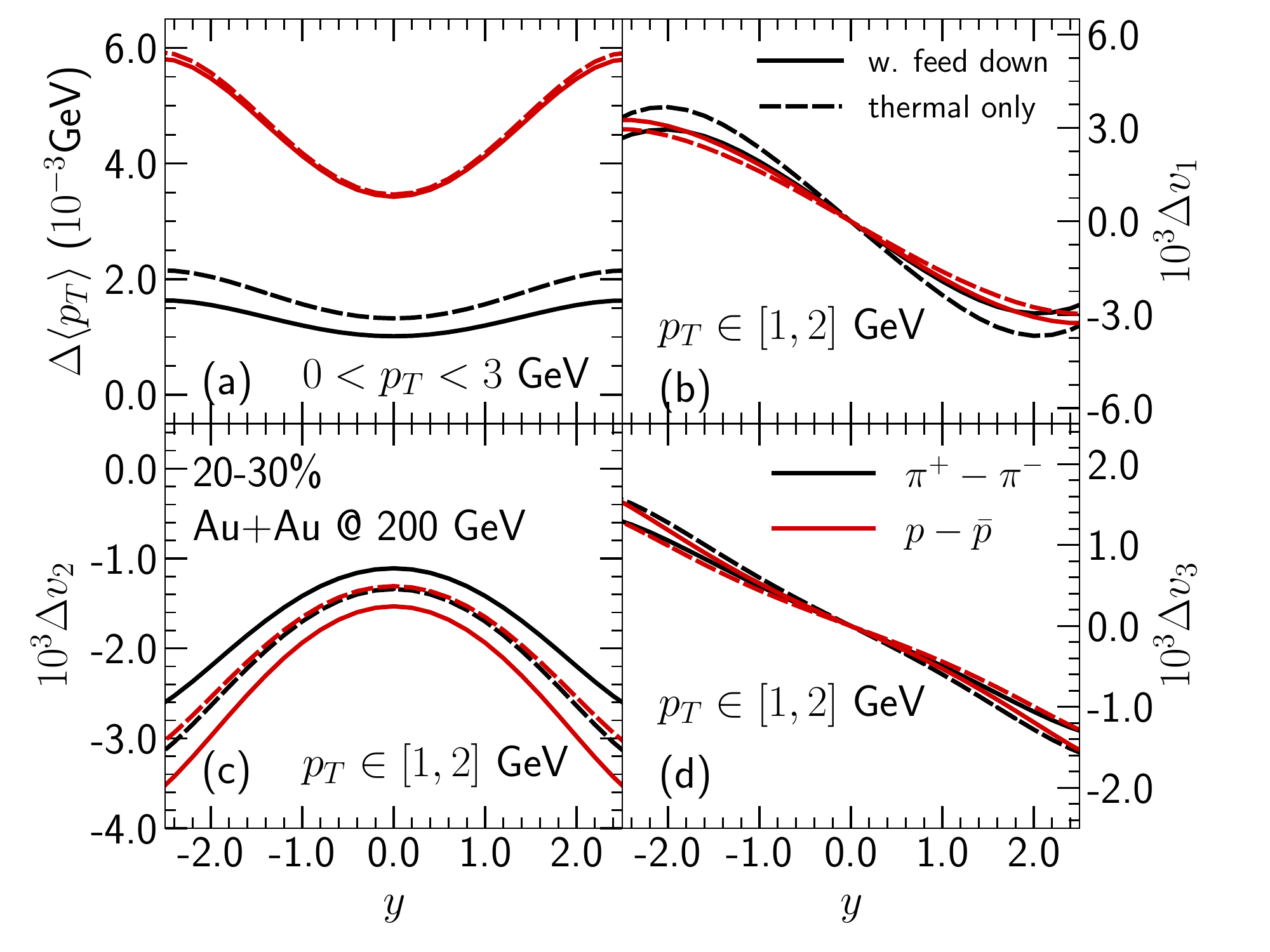}   
  \caption{The solid curves include the contributions to the electromagnetically induced
  charge-dependent flow observables of pions and protons produced after freeze-out by  resonance decay, often referred to as resonance feed-down contributions. In the dashed curves, pions and protons produced from resonance feed-down are left out.
}
  \label{fig6}
\end{figure*}

Finally, in Fig.~\ref{fig6}, we investigate the contribution of resonance decays to the electromagnetically induced charge-dependent contributions to flow observables that we have computed. These contributions are included in all our calculations with the exception of those shown as the dashed lines in Fig.~\ref{fig6}, where we include only the hadrons produced directly at freezeout, leaving out those produced later as resonances decay.  
We see that the feed-down contribution from resonance decays does not significantly dilute the effects we are interested in. To the contrary, the magnitudes of the $\Delta v_n$ for protons are slightly increased by feed-down effects, in particular the significant contribution to the final proton yield coming from the decay of the $\Delta^{++}$~\cite{Qiu:2012tm}. Because the $\Delta^{++}$ resonance carries 2 units of the charge, its electromagnetically induced drift velocity is larger than those of protons.

This concludes the presentation of our central results.  In the remainder of this Section, in two  subsections we shall present a qualitative argument for why $\Delta v_1$ is as small as it is, and then take a brief look at how  our results depend on the value of two important material properties of the plasma, namely the drag coefficient and the electrical conductivity.

\subsection{A qualitative argument for the smallness of $\Delta v_1$}
\label{sec::cancel}

As we have seen, the net effect on $\Delta v_1$ of the various contributions to the electric field turns out to be rather small in magnitude. 
This is because even though the contributions  $\vec{E}_C+\vec{E}_F$  and $\vec{E}_L$ with opposite sign, shown separately in Fig.~\ref{figExeta13}, are each relatively large in magnitude they cancel each other almost precisely. This leaves only a small net contribution that generates the charge-odd contributions to the odd flow harmonics that we have computed, $\Delta v_1$ and $\Delta v_3$. We see in Fig.~\ref{figExeta13} that this cancellation becomes more and more complete at larger $\eta_s$.  In this subsection we provide a qualitative argument for this near-cancellation and explain why the cancellation becomes more complete at larger $\eta_s$.

One can find an expression for the total Faraday+Coulomb electric field $\vec{E}_{F+C} \equiv \vec{E}_F + \vec{E}_C$ by solving the Maxwell equations sourced by the spectator (and participant\footnote{To a very good approximation, one can in fact ignore the participant contribution \cite{Gursoy:2014aka}.}) charges. In general this determines both the electric and the magnetic fields in terms of the sources. However, we only need to express $\vec{E}_{F+C}$ in terms of $\vec{B}$ for the argument. In particular, we are interested in the $x$ component of this field as shown in Fig.~\ref{figschema}. This is given by solving Faraday's law $\nabla \times \vec{E}_{F+C} = -\partial \vec{B}/\partial t$ to obtain $E_{F+C,x} = B_y \coth(Y_0-\eta_s)$, where $Y_0$ is the rapidity of the beam and $\eta_s$ is the spacetime rapidity. 
Since for both RHIC and LHC we have $Y_0\gg \eta_s$, one can safely ignore the $\eta_s$-dependence everywhere in the plasma, finding $E_{F+C,x} \approx B_y \coth(Y_0)$. For the same reason, as $Y_0\gg 1$, one can further approximate $E_{F+C,x} \approx B_y$ everywhere in the plasma. The effect of this electric field on the drift velocity of the plasma charges  is found by solving the null-force equation (\ref{eq9}) by boosting it to the local fluid rest frame in a given unit cell in the plasma. 
This gives the contribution $E_{F+C,x}^{\rm\, lrf} \approx \gamma(u) B_y$ where $\gamma(u) $ is the Lorentz gamma factor of the plasma moving with velocity $u$. On the other hand, the $x$-component of the Lorentz contribution to the force in the local fluid rest frame is to a very good approximation given by $E_{L,x}^{\,\rm lrf} = - \gamma(u) u_z B_y$, where $u_z=\tanh\eta_s$ is the $z$-component of the background flow velocity. 
As is clear from Fig.~\ref{figschema}, the directed flow coefficient $v_1$ receives its largest  contribution from sufficiently large $\eta_s$ where $u_z\approx 1$. 
We now see that in the regime $ 2  \lesssim \eta_s \ll Y_0$ there is an almost perfect cancellation between  $E_{L,x}^{\rm\, lrf}$ and $E_{F+C,x}^{\rm\, lrf}$, with $E_{L,x}^{\rm\, lrf}$ slightly smaller on account of the fact that $u_z$ is slightly smaller than 1.
This means that the main contribution to $\Delta v_1$ should come from the mid-rapidity region where the cancellation is only partial as illustrated in Fig.~\ref{figExeta13}, meaning that $\Delta v_1$ is bound to be small in magnitude.

\subsection{Parameter dependence of the results}
\label{sec::parameters}

\begin{figure*}[t!]
  \includegraphics[width=0.8\linewidth]{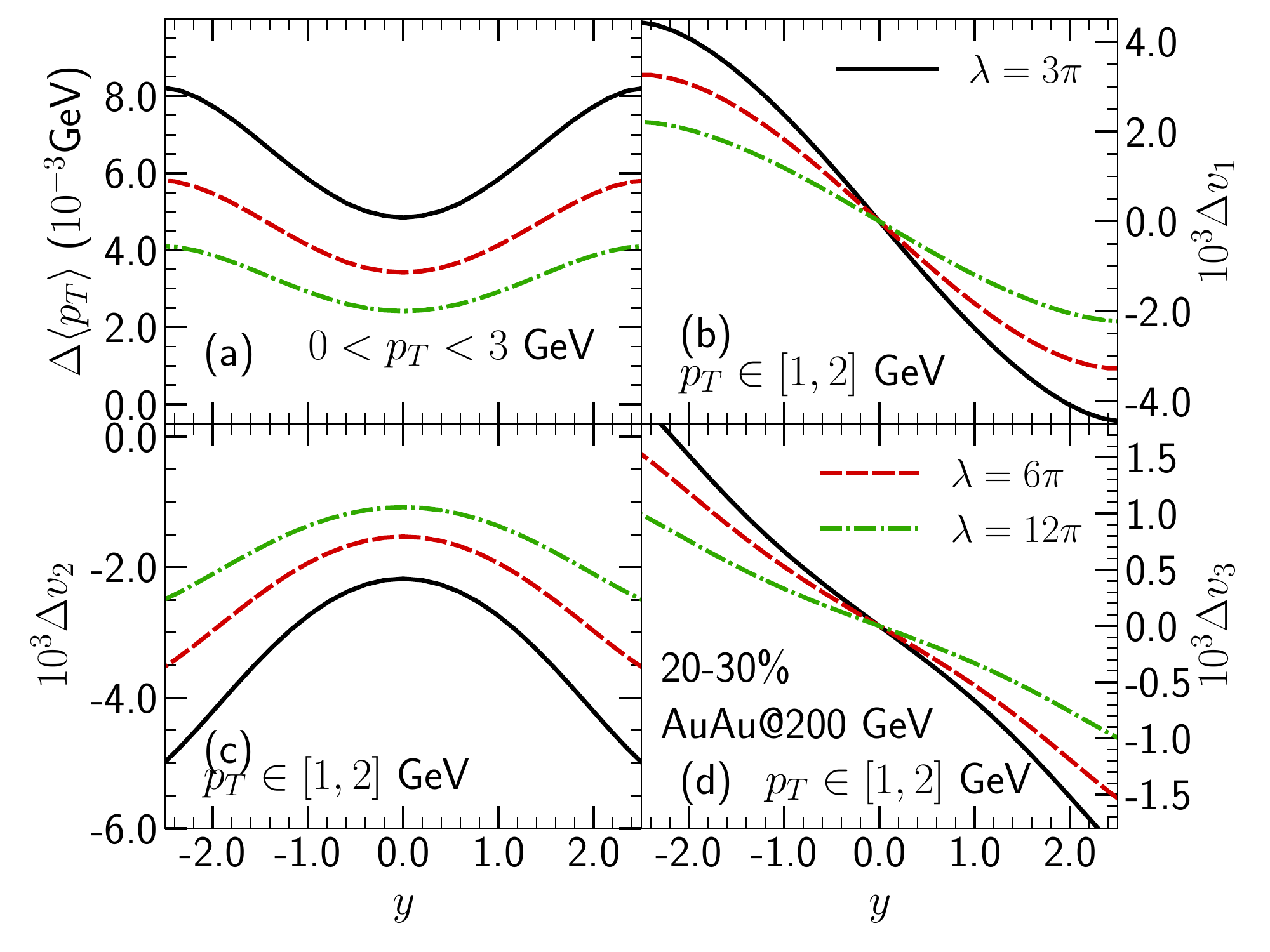}   
  \caption{The dependence of the electromagnetically induced differences between the flow of protons and antiprotons on the choice of the drag coefficient $\mu m$ defined in Eq.~(\ref{eq10}). Elsewhere in this paper, we fix $\mu m$ by choosing the 't-Hooft coupling in Eq.~(\ref{eq10}) to be $6\pi$. Here we explore the consequences of varying this parameter by factors of 2 and 1/2, thus varying $\mu m$ by factors of $\sqrt{2}$ and $1/\sqrt{2}$.    }
  \label{fig7}
\end{figure*}

Throughout this paper, we have chosen fixed values for the two important material parameters that govern the magnitude of the electromagnetically induced contributions to flow observables, namely the drag coefficient $\mu m$ defined in Eq.~(\ref{eq10}) and the electrical conductivity $\sigma$.  Here we explore the consequences of choosing different values for these two parameters.

In Fig.~\ref{fig7}, we study the effect of varying the drag coefficient $\mu m$ 
on the the magnitude of the electromagnetically induced differences between the flow of
protons and antiprotons. We change the value of the drag coefficient in Eq.~(\ref{eq10}) by choosing different values of the 't Hooft coupling $\lambda$. (The consequences of varying $\mu m$ 
for the differences between 
the flow of $\pi^+$ and $\pi^-$ are similar, although the magnitude of the $\Delta v_n$'s is less for pions than for protons.)
We see in Fig.~\ref{fig7} that all of the charge-dependent contributions to the flow  that are induced by electromagnetic fields become larger when the drag coefficient $\mu m$ becomes smaller, as at weaker coupling.  
This is because  the induced drift velocity $v_\mathrm{drift}^\mathrm{\rm\, lrf}$ in equation (\ref{eq9}) is larger when the  drag coefficient $\mu m$ is smaller.
Since throughout the paper we have used a value of $\mu m$ that is motivated
by analyses of drag forces in strongly coupled plasma, meaning that we may have overestimated $\mu m$, it is possible that in so doing we have underestimated the magnitude of the charge-odd electromagnetically induced contributions to flow observables.

\begin{figure*}[t!]
  \includegraphics[width=0.8\linewidth]{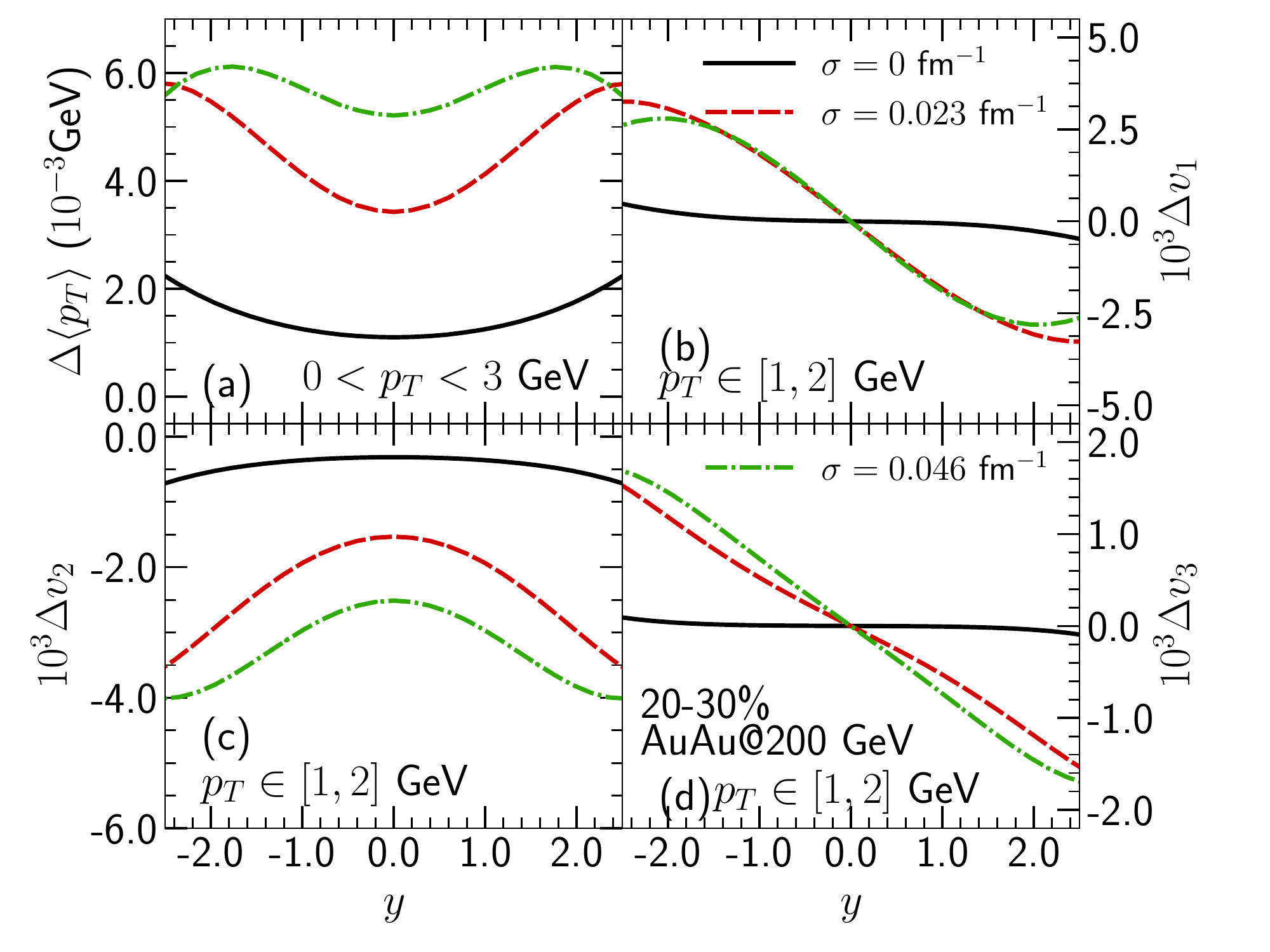}   
  \caption{The dependence of the electromagnetically induced differences between the flow of protons and antiprotons on the choice the electrical conductivity $\sigma$ in the Maxwell equations (\ref{eq7}) and (\ref{eq8}). 
    }
  \label{fig8}
\end{figure*}

In Fig.~\ref{fig8}, we study the effect of varying the electrical conductivity $\sigma$ on the 
magnitude of the electromagnetically induced differences between the flow of protons and antiprotons.
Note that, throughout, we are treating $\mu m$ and $\sigma$ as constants, neglecting their temperature dependence.  This is appropriate for $\mu m$, since what matters in our analysis is the value of $\mu m$ at the freezeout temperature.  However, $\sigma$ matters throughout our analysis since it governs how fast the magnetic fields sourced initially by the spectator nucleons decay away.  The value of $\sigma$ that we have used throughout the rest of this paper is reasonable for quark-gluon plasma with a temperature $T\sim 250$~MeV, as we discussed in Section~\ref{sec::model}.  
In a more complete analysis, $\sigma$ should depend on the plasma temperature and hence should vary in space and time.  We leave a full-fledged magnetohydrodynamic study like this to the future. Here, in order to get a sense of the sensitivity of our results to the choice that we have made for  $\sigma$ we explore the consequences for our results of doubling $\sigma$, and of setting $\sigma=0$.

The electromagnetically induced charge-odd contributions to the flow observables 
$\Delta \langle p_T \rangle$ and $\Delta v_2$ increase in magnitude if the value of $\sigma$ is increased.  
This is because  the  magnetic fields in the plasma decay more slowly 
when $\sigma$ is large~\cite{Gursoy:2014aka}. And,
a larger electromagnetic field in the local fluid rest frame at the freezeout surfaces induces
a larger drift velocity which drives the opposite contribution to proton and antiproton flow observables.
We see, however, that the increase in the charge-odd, rapidity-odd, odd $\Delta v_n$'s with increasing $\sigma$ is very small, suggesting a robustness in our calculation of their magnitudes.
This would need to be confirmed via a full magnetohydrodynamical calculation in future.
Since $\Delta \langle p_T \rangle$ and the even $\Delta v_n$'s are to a significant degree driven
by Coulomb fields, it makes sense that they are closer to proportional to $\sigma$: increasing $\sigma$ means that a given Coulomb field pushes a larger current, and it is the current in the plasma that leads to the charge-odd contributions to flow observables.
Although not physically relevant, it is also interesting to check the consequences of 
setting $\sigma=0$.  
What remains are small but nonzero contributions to $\Delta \langle p_T \rangle$ and the $\Delta v_n$.
With $\sigma=0$ the electric fields do not have any effects during the Maxwell evolution;
the small remnant fields at freezeout are responsible for these effects.

\section{Discussion and Outlook}
\label{sec::discussion}

We have described the effects of electric and magnetic fields on the flow of charged hadrons in non-central heavy ion collisions by using a realistic hydrodynamic evolution within the {\tt iEBE-VISHNU} framework.
The electromagnetic fields are generated mostly by the spectator ions.    These fields induce
a rapidity-odd contribution to $\Delta v_1$ and $\Delta v_3$ of charged particles, 
namely the difference between $v_1$ (and $v_3$) for positively and negatively charged particles.
Three different effects contribute: the Coulomb field of the spectator ions, the Lorentz force due to
the magnetic field sourced by the spectator ions, and the electromotive force induced by Faraday's law as that magnetic field decreases.
The $\Delta v_1$ and $\Delta v_3$ in sum arise from a competition between the Faraday and Coulomb effects, which point in the same direction, and the Lorentz force, which points in the opposite direction. 
These effects also induce a rapidity-even contribution to $\Delta \langle p_T \rangle$ and $\Delta v_2$, as does the Coulomb field sourced by the charge within the plasma itself, deposited therein by the participant ions.
We have estimated the magnitude of all of these effects for pions and protons produced in heavy ion collisions with varying centrality at RHIC and LHC energies. Our results motivate the experimental measurement of these quantities with the goal of seeing observable consequences of 
the strong early time magnetic and electric fields expected in ultrarelativistic heavy ion collisions. 

In our calculations, we have treated the electrodynamics of the charged matter in the plasma in a perturbative fashion, added on top of the background flow, rather than attempting a full-fledged magnetohydrodynamical calculation.  The smallness of the effects that we find supports this approach.
However, we caution that we have made various important assumptions that simplify our calculations: (i) we treat the two key properties of the medium that enter our calculation, the electrical conductivity  $\sigma$ and the drag coefficient $\mu m$, as if they are both constants 
even though we know that both are temperature-dependent and hence in reality must vary in both space and time within the droplet of plasma produced in a heavy ion collision; (ii) we neglect event-by-event fluctuations in the shape of the collision zone; (iii) rather than full-fledged magnetohydrodynamics, we follow a perturbative calculation where we neglect backreaction of various types, including the rearrangement of the net charge in response to the electromagnetic fields; (iv) we assume that the force-balance equation (\ref{eq9}) holds at any time and at any point on the plasma, meaning that we assume that the plasma equilibrates immediately by balancing the electromagnetic forces against drag. 
As we shall discuss in turn, relaxing these assumptions could have interesting consequences, and is worthy of future investigation. But, relaxing any of these assumptions would result in a substantially more
challenging calculation.

Relaxing (i) necessitates solving the Maxwell equations on a medium with time- and space-dependent parameters, which would result in a more complicated profile for the electromagnetic fields. 
We expect that this would modify our results in a quantitative 
manner without altering main qualitative findings. 
We have tried to choose
a value for $\sigma$ corresponding roughly to a time average over the lifetime of the plasma
and a value of $\mu m$ corresponding roughly to its value at freezeout, which is where it
is relevant to our analysis.  The values of each
could be revisited, of course, but our investigation in Section~\ref{sec::parameters}  indicates that this would not affect any
qualitative results.

Relaxing (ii), which is to say adding event-by-event fluctuations in the initial conditions for
the hydrodynamic evolution of the matter produced in the collision zone, as well as for
the distribution of spectator charges,
would have quite significant effects on the values of the charge-averaged $\langle p_T \rangle$
and $v_n$'s, for example introducing nonzero  $v_1$ and $v_3$.
Solving the Maxwell equations on such a medium would of course be much more complicated.  Furthermore
we expect that consequences would appear in all four of the electromagnetic effects
that we have analysed (the Faraday $\vec E_F$, the Lorentz $\vec E_L$, the
Coulomb field of the spectators $\vec E_C$ and the Coulomb field of the plasma $\vec E_P$) resulting in each
contributing at some level to each of the four observables that we have analysed ($\Delta \langle p_T \rangle$, $\Delta v_1$, $\Delta v_2$ and $\Delta v_3$). However, 
we expect that the electromagnetically induced contributions that we have found using a smooth hydrodynamic background without fluctuations, and whose magnitudes we have estimated, will remain the largest contributions.

Relaxing assumption (iii) may bring new effects and, as we shall explain, could 
potentially flip the sign of the odd flow coefficients $\Delta v_1$ and $\Delta v_3$. 
One particular physical effect that we neglect is the  shorting, or partial shorting, 
of the Coulomb electric fields
in the plasma, both the $\vec E_C$ sourced by the spectators and the $\vec E_P$ sourced
by the plasma itself.
These Coulomb fields will push charges in the plasma to rearrange in a way that
reduces the electric field within the conducting plasma.  
We have neglected this, and all, back reaction in our calculation. However, although it
would require a fully dynamical calculation of the currents and electric and magnetic
fields to estimate its extent,
some degree of shorting must occur.
There may, in fact, be experimental evidence of this effect: $\Delta v_2$ for pions 
has been measured in RHIC collisions with 30-40\% centrality and collision energy $\sqrt{s}=200$~AGeV by the STAR collaboration~\cite{Adamczyk:2015eqo}, and although it turns out to be negative as our calculations 
predict it is substantially smaller in magnitude than what we find.  Because
there are other effects (unrelated to Coulomb fields) that can contribute to $\Delta v_2$ and that are known to contribute significantly to $\Delta v_2$ in lower energy collisions~\cite{Xu:2012gf,Steinheimer:2012bn,Song:2012cd,Adamczyk:2013gv,Xu:2013sta,Adamczyk:2015fum,Xu:2016ihu}, it would take substantially more analysis
than we have done to use the experimentally measured results for $\Delta v_2$ to
constrain the magnitude of $\vec E_C$ and $\vec E_P$ quantitatively. However, it does
seem likely that, due to back reaction, they have been at least partially
shorted, making them weaker in reality than in our calculation.

The likely reduction in the magnitude of $\vec E_C$, in turn, has implications for the odd $\Delta v_n$'s.  Recall that they arise
from the sum of three effects, in which there is a near cancellation between
$\vec E_F+\vec E_C$ and $\vec E_L$, which point in opposite directions.  
The sign of the rapidity-odd $\Delta v_1$ and $\Delta v_3$
that we have found in our calculation corresponds to $| \vec E_F + \vec E_C |$ being
slightly greater than $| \vec E_L |$.  If $|\vec E_C |$ is in reality smaller than in our
calculation, this could easily flip the sign of $\Delta v_1$ and $\Delta v_3$.
In this context, it is quite interesting that a preliminary analysis of ALICE data~\cite{Margutti:2017lup}
indicates a measured value of $\Delta v_1$ for charged particles in LHC heavy ion collisions with 5\%-40\% centrality and collision energy $\sqrt{s}=5.02$~ATeV 
that is indeed rapidity-odd and is comparable in magnitude to the pion $\Delta v_1$ for collisions with this energy that we have found in Fig.10, but is opposite in sign.

Finally, let us consider relaxing our assumption (iv).
This corresponds to considering a more general version of (\ref{eq9}) with a non-vanishing acceleration on the right-hand side.
The drift velocity that would be obtained in such a calculation would decay to the one that we have found by solving the force-balance equation (\ref{eq9}) exponentially, with an exponent controlled by the drag coefficient $\mu$.
Thus, for very large $\mu$ we do not expect any significant deviation from our results. 
However, at a conceptual level relaxing assumption (iv) would change our calculation significantly,
since it is only by making assumption (iv) that we are able to do a calculation in which $\mu$ enters only through the value of $\mu m$ at freezeout.  If we relax assumption (iv), the actual drift velocity would always be lagging behind the value obtained by solving (\ref{eq9}), and determining the drift velocity at freezeout would, in principle, retain a memory of the history of the time evolution of $\mu$. If we use the estimate (\ref{eq10}) for $\mu$ and focus only on light quarks, and hence pions and protons, as we have done we do not expect that relaxing assumption (iv) would have a qualitative effect on our results.  However, $\mu$ may in reality not be as large as that in (\ref{eq10}) at freezeout. And, furthermore, it is also very interesting to extend our considerations to consider heavy charm quarks, as in Ref.~\cite{Das:2016cwd}. The charm quarks receive a substantial initial kick from the strong early time  magnetic~\cite{Das:2016cwd} and electric fields, and because they are heavy  $\mu$ may not be large enough to slow them down and bring them into alignment with the small drift velocity that  (\ref{eq9}) predicts for heavy quarks.
Hence, consideration of heavy quarks requires relaxing our assumption (iv) in a way that alters our conclusions significantly, and indeed the authors of Ref.~\cite{Das:2016cwd} find a substantially larger $\Delta v_1$ for mesons containing charm quarks than the $\Delta v_1$ that we find for pions and protons. These considerations motivate the (challenging) experimental measurement of $\Delta v_1$ for $D$ mesons.

\begin{acknowledgements}
This work was supported in part by the Netherlands Organisation for Scientific Research (NWO) under VIDI grant 680-47-518, the Delta Institute for Theoretical Physics (D-ITP) funded by the Dutch Ministry of Education, Culture and Science (OCW), the Scientific and Technological Research Council of Turkey (TUBITAK), the Office of Nuclear Physics of the U.S.~Department of Energy under Contract Numbers  DE-SC0011090, DE-FG-88ER40388 and DE-AC02-98CH10886, and the Natural Sciences and Engineering Research Council of Canada. KR gratefully acknowledges the hospitality of the CERN Theory Group. CS gratefully acknowledges a Goldhaber Distinguished Fellowship from Brookhaven Science Associates. Computations were made in part on the supercomputer Guillimin from McGill University, managed by Calcul Qu\'ebec and Compute Canada. The operation of this supercomputer is funded by the Canada Foundation for Innovation (CFI), NanoQu\'ebec, RMGA and the Fonds de recherche du Qu\'ebec -- Nature et technologies (FRQ-NT). UG is grateful for the hospitality of the Bo\~gazi\c ci University and the Mimar Sinan University in Istanbul. We gratefully acknowledge helpful discussions with Gang Chen, Ulrich Heinz, Jacopo Margutti, Raimond Snellings, Sergey Voloshin and Fuqiang Wang. 
\end{acknowledgements}

\bibliography{references}

\end{document}